\documentclass[aps,prb,superscriptaddress,showpacs,reprint]{revtex4-1}
\usepackage{graphicx}
\usepackage{subfigure}
\usepackage{caption}
\usepackage{epstopdf}
\usepackage{epsfig}

\usepackage{amssymb}
\usepackage{amsthm}

\begin{document}

%\preprint{}

\title{Assessing thermal spike model of swift heavy ion-matter interaction via Pd$_{1-x}$Ni$_x$/Si interface mixing}


\author{Paramita Patra}
\email[]{patro.paro@phy.iitkgp.ernet.in}
%\thanks{}
\affiliation{Department of Physics, Indian Institute of Technology Kharagpur, Kharagpur - 721302, INDIA}

\author{S. A. Khan}
%\email[]{sunit@chem.iitkgp.ernet.in }
%\thanks{}
\affiliation{Inter-University Accelerator Centre, Aruna Asaf Ali
Marg, New Delhi - 110067, INDIA}

\author{M. Bala}
%\email[]{sunit@chem.iitkgp.ernet.in }
%\thanks{}
\affiliation{Department of Physics and Astrophysics, University of
Delhi, New Delhi - 110007, INDIA}

\author{D. K. Avasthi}
%\email[]{sunit@chem.iitkgp.ernet.in }
%\thanks{}
\affiliation{Amity Institute of Nanotechnology, Amity University,
Sector 125, Noida - 201313, INDIA}

\author{S. K. Srivastava}
\email[]{sanjeev@phy.iitkgp.ernet.in}
%\thanks{}
\affiliation{Department of Physics, Indian Institute of Technology
Kharagpur, Kharagpur-721302, INDIA}


\date{\today}

\begin{abstract}

Thermal spike model (TSM) is presently a widely accepted mechanism of swift heavy ion (SHI) - matter interaction. It provides explanation to various SHI induced effects including mixing across interfaces. The model involves electron-phonon (e-p) coupling to predict the evolution of lattice temperature with time. SHI mixing is considered to be a result of diffusion in transient molten state thus achieved. In this work, we assess this conception primarily via tuning the e-p coupling strength by taking a series Pd$_{1-x}$Ni$_x$ of a completely solid soluble binary, and then observing 100 MeV Au ion induced mixing across Pd$_{1-x}$Ni$_x$/Si interfaces. The extent of mixing has been parametrised by the irradiation induced change $\Delta \sigma^2$ in variances of Pd and Ni depth profiles derived from X-ray photoelectron spectroscopy. The $x$-dependence of $\Delta \sigma^2$ follows a curve that is concave upward with a prominent minimum. Theoretically, e-p coupling strength determined using density functional theory has been used to solve the equations appropriate to TSM, and then an equivalent quantity L$^2$ proportional to $\Delta \sigma^2$ has been calculated. L$^2$, however, increases monotonically with $x$ without any minimum, bringing out a convincing disparity between experiment and theory. Perhaps some mechanisms more than the TSM plus the transient molten state diffusion are operative, which can not be foreseen at this point of time.

\end{abstract}

%\noindent{\it Keywords}: XXXXXX, YYYYYYYY, ZZZZZZZZZ \\
%\submitto{\NT}
\maketitle

\section{Introduction}

While interacting with a solid, swift heavy ions (SHI's) are known
to transfer a large amount of energy predominantly in the
electronic subsystem of the solid.\cite{Northcliffe} At first
hand, therefore, SHI's are not anticipated to cause any atomic
displacements in the solid they interact with. However,
irradiation of solids with SHI's has often been found to result in
atomic displacements, which may cause, among various effects,
mixing across interfaces in layered materials.\cite{Bsch03,CLT15,SKS14} There certainly has to be a mechanism
by which the electronic energy, deposited by the SHI in the solid,
gets transferred to the lattice atoms to cause such atomic
displacements. At present, there are essentially two established
models, viz., the Coulomb spike model (CSM) and the thermal spike
model (TSM), to explain such atomic displacements.\cite{ZW94,MTOU02,AG01,SKS05} According to the CSM, a SHI, while
passing through a solid material, ionizes the
material in a cylindrical region around its path.  The consequent
strong collective electrostatic repulsion amongst the positive ions
in the ionized zone leads to violent atomic displacements,
resulting ultimately into a modified material in a cylindrical
so-called ion track. This model, however, lacks applicability in
metals, where the high mobility of conduction electrons leads to
neutralization of charges much before the Coulomb explosion could
occur. The TSM, on the other hand, assumes that the energy
deposited initially in the electronic subsystem in a time scale of
$10^{-15} - 10^{-14}$ s gets subsequently transferred to the
lattice subsystem via electron-phonon (e-p) coupling in $\sim
10^{-13} -10^{-12}$ s. This results in a rapid rise in the lattice
temperature (up to $\sim 10^{4}$ K) in a cylindrical zone of
typically a few nm radius. In certain conditions a molten state is
created along the ion track for a $\sim 10^{-12} - 10^{-11}$ s
duration and is quenched rapidly (at a rate of $\sim 10^{14}$ K/s),
freezing the molten modified state of the cylindrical zone. The
modified frozen cylindrical zone thus formed is conventionally
known as a latent tack. If such melting happens across an
interface between two materials, the atoms on the two sides
interdiffuse while in the molten state, giving rise to mixing
across the interface.\cite{SKS05} The mechanism suggests that the
TSM must be applicable in metals, semiconductors and insulators
alike. The model has acquired a wide acceptance in course of time.

The TSM is mathematically described by the following two coupled
partial differential equations, which are basically the
constituents of the so-called two temperature model (TTM) and
govern the diffusion of the energy brought in by the ion into the
electronic and lattice subsystems:\cite{ZW94,CDu93}

%\begin{center}
\begin{math}
 C_e(T_e)\frac{\partial T_e}{\partial t}=\nabla.(K_e(T_e)\nabla T_e)-(T_e-T_l) G(T_e)+A_e(r,t)
%\label{seqn1}
\end{math}
and \noindent
\begin{equation}
 C_l(T_l)\frac{\partial T_l}{\partial t}=\nabla.(K_l(T_l)\nabla T_l)+(T_e-T_l) G(T_e).
%\label{seqn2}
\end{equation}
%\end{center}
\noindent Here, $C_e$, $C_l$ and $K_e$, $K_l$ stand for specific
heats and thermal conductivities of the electronic and lattice
subsystems, and $T_e$ and $T_l$ are the electronic and lattice
temperatures, respectively. $G(T_e)$ is the electronic temperature
dependent e-p coupling strength, and $A_e (r,t)$ is the energy
density per unit time supplied by the incident ions to the
electronic system at time $t$ and at radius $r$ from the ion path
in such a way that the integral $\int\int 2\pi r A_e(r,t) \,dr
\,dt$ is equal to the electronic energy loss $S_e$, defined as the
energy deposited by the ions in the electronic subsystem per unit
length travelled in the solid.

A direct experimental proof of the validity of the TSM has
hitherto not been possible because of the extremely short time
scales involved. A number of ion fluence dependent SHI induced
effects observed experimentally have been used to coarsely derive
the latent track radii, which have been found comparable with
those calculated roughly from the mathematical equations
pertaining to the TSM.\cite{SKS05} These provide a highly
indirect and very crude indication of the occurrence of SHI
induced processes as hypothesized in the TSM. In all these
reports, the free electron theory of metals, which predicts a
parabolic density of electron states (eDOS), has been used to
determine the electronic part of the thermophysical parameters,
viz. $C_e$, $K_e$ and $G(T_e)$,\cite{ZW94,MTou96} to be used in
the TTM equations. Accordingly, $C_e$ is given by\cite{ZW94}

\begin{equation}
 C_e(T_e) = {{\pi^2 g(\epsilon_F) k_B^2}\over{2 \epsilon_F}}T_e,
\end{equation}
$K_e$ is related with $C_e$ via electronic thermal diffusivity
$D_e(T_e)$ by a relation
\begin{equation}
 K_e(T_e) = C_e(T_e)D_e(T_e),
\end{equation}
and $G(T_e)$ is determined using
\begin{equation}
 G_e(T_e) = {{\pi^4 [g(\epsilon_F) k_B v_s]}^2\over{18 K_e(T_e)}}.
\end{equation}
Here, $g(\epsilon_F)$ is the eDOS at the Fermi energy
$\epsilon_F$, and $k_B$ is the Boltzmann constant. The phonon
contribution to the e-p coupling strength appears in the form of
the speed $v_s$ of sound in the solid. However, a couple of
reports on 120 MeV Au induced mixing in Si/M/Si (M = V, Fe, Co,
Mn, Nb) layered structures \cite{brc06,kd04} suggested that the
relatively more localized $d$-electrons, which bring in features
to the eDOS over the parabolic background, also have influence on
the efficiency of SHI mixing. This necessitates the consideration
of exact electron density of states $g(\epsilon)$ as a function of
energy $\epsilon$, computable using $first-principles$ density
functional theory (DFT), to derive the thermophysical quantities
required for the TTM. The following forms of $C_e$ and $G(T_e)$,
as reported by Lin $et \,al.$,\cite{zl08} would be more
appropriate in this scenario:
\begin{equation}
C_e(T_e)=\int_{-\infty}^{\infty} g(\epsilon)\Big[\frac{\partial f(\epsilon,T_e)}{\partial T_e}\Big]\epsilon \,d\epsilon
\end{equation}
and
\begin{equation}
G(T_e)=\frac{h k_B \lambda \left\langle
\omega^2\right\rangle}{2g(\epsilon_F)}\int_{-\infty}^{\infty}
g^2(\epsilon)\Big[-\frac{\partial f(\epsilon,T_e)}{\partial
\epsilon}\Big] d\epsilon,
\end{equation}
\noindent where $h$ is the Planck's constant and $f(\epsilon,T_e)$
is the Fermi-Dirac distribution function given by $f(\epsilon,
T_e)=1/[1+{\rm exp}\{(\epsilon-\epsilon_F)/k_BT_e\}]$. The
electron band mass enhancement factor $\lambda$\cite{DAP76,GGRIM76} and the second moment $\left\langle
\omega^2\right\rangle$ of the phonon spectrum\cite{AKGi86} can be
obtained from $ab \,\,initio$ phonon bandstructure calculations.
Lin $et \,al.$,\cite{zl08} this way, have calculated electron
temperature dependent electronic specific heats and electron phonon
coupling strengths for a number of noble and transition metals, and
have reported a substantial difference between the free-electron
and full eDOS values. In a recent work,\cite{PP16} we have shown
how slight variations of $g(\epsilon)$ for different orientations
of a thin $\rm{Bi_2Te_3}$ slab result into different $C_e(T_e)$
and $G(T_e)$ curves.

If the TSM is valid, the use of equations (6) and (7), instead of
(3) and (5), in the TTM equations ought to improve the predicting
ability of the model for getting an outcome of a SHI-matter
interaction experiment. To enact this, conducting a series of
SHI-matter interaction experiments, e.g. SHI driven interface
mixing across a number of thin film/substrate interfaces, with one
kind of substrate and different kinds of thin films of differing
$C_e$ and $G$ values, would be helpful. As far as the rest of TTM
parameters, viz. $K_e$, $C_l$ and $K_l$, are concerned, they
could, to an appreciable extent, be predictable or obtainable from
literature for each film. One possibility could be taking a series
of M/Si interfaces with different metals M so that the eDOS and
the resultant $C_e$ and $G$ values could be calculated for each M.
However, the $K_e$, $C_l$ and $K_l$ values might be arbitrarily
different for different M, a case which is obviously undesirable.
An appropriate choice for M would be to take thin films
A$_{1-x}$B$_x$ of a complete solid-soluble binary metal system
with $0 \leq x \leq 1$. For such a series, $g(\epsilon)$, $C_e$
and $G$ would be easily computable for each $x$.
Furthermore, the $K_e$, $C_l$ and $K_l$ values will have a smooth
(to the first approximation linear) variation with $x$.\cite{Sch98} One such system is Pd$_{1-x}$Ni$_x$ binary alloy
system, which forms a complete solid solution throughout the whole
composition range without any change of the crystal structure, as
depicted by their equilibrium phase diagram.\cite{dd}

The present work aims at convincingly assessing the thermal spike
model by (i) experimental determination of the $x$-variation of
efficiency of SHI driven mixing of Pd and Ni in Si via 100 MeV Au
irradiation of Pd$_{1-x}$Ni$_x$/Si system, (ii) computation of
$x$-variation of $C_e$ and $G$ using DFT, and then use of the TTM
equations to qualitatively estimate the expected $x$-variation of
extent of mixing, and (iii) a comparison between the experimental
and computational results. Any slight variation in the
computationally predicted efficiency of mixing should be
observable also in the experimental results if the TSM is indeed
the mechanism of SHI matter interaction. However, neither Pd nor
Ni is known to be mixed with Si by SHI's; Pd/Si or Ni/Si mixing
has only been reported to be induced by low energy ions,\cite{BYT79,JD92} where elastic collisions are responsible for
the process. In the present work also, Pd/Si or Ni/Si mixing in
the Pd$_{1-x}$Ni$_x$/Si system has not been observed as an
immediate effect of the irradiation; it is rather the Ar$^+$ ion
sputtering process in the subsequent X-ray photoelectron
spectroscopy (XPS) depth profiling that augments the effects of
SHI's and leads to observable mixing. Since the sputtering
conditions are the same for all the samples, any $x$-variation of
extent of mixing should have indirectly been driven by the SHI
irradiation. Such a combination of SHI irradiation and XPS depth
profiling to enable one to observe SHI effects in the form of
interface mixing has hitherto not been reported.

\section{Experimental and computational details}

Prior to depositing the Pd$_{1-x}$Ni$_x$ ($x \neq 0, 1$) alloy
thin films on Si substrates, the alloys were first prepared by Ar
arc melting. Palladium wire of 99.9\% purity and nickel foil of
99.994\% purity were melted together to prepare the alloys. The
alloyed ingots were flipped and remelted to improve the
homogeneity. Subsequently, Pd and Ni metals, and two compositions
$x$ = 0.40 and 0.78 of Pd$_{1-x}$Ni$_x$ alloys were deposited onto
pre-cleaned Si substrates by electron beam evaporation. The
pressure during deposition was $\sim 1.7$ $\times$ $10^{-7}$ torr,
and the deposition rate ranged from 0.1 to 0.3 \AA/s. The
thicknesses of the four Pd$_{1-x}$Ni$_x$ ($x$ = 0, 0.40, 0.78 and
1) films were in 25 - 40 nm range, as determined from Rutherford
backscattering spectra (RBS) and XPS depth profiles to be
discussed in the following sections. For all the samples, 1 cm
$\times$ 1 cm pieces were taken out for irradiation by 100 MeV Au
ions each at 1$\times$ $10^{14}$ ions/cm$^{2}$ fluence using the
15 UD Pelletron accelerator at Inter University Accelerator Centre
(IUAC), New Delhi. The electronic ($S_e$) and nuclear ($S_n$)
energy losses of the ions in Pd are 34.10 keV/nm and 0.09 keV/nm,
respectively, as calculated from the SRIM software.\cite{JFZei85}
The two values for Ni are 32.40 keV/nm and 0.08 keV/nm,
respectively. Thus, the electronic energy losses are dominant and
hence the condition is relevant to the present study. The pristine
and irradiated samples were characterized by X-ray diffraction
(XRD), RBS, and XPS depth profiling. XRD patterns for all the
samples were recorded using Cu K$_\alpha$ radiation from a Philips
X'Pert MRD X-ray diffractometer in the 2$\theta$ range of
$20^{\circ} \- 80^{\circ}$. The RBS measurements were performed
using 2 MeV He$^+$ ions from the PARAS facility of IUAC, and were
undertaken to determine the thickness and composition of the
samples. The XPS depth profiles were carried out using a PHI 5000
Versaprobe II machine under DST-FIST scheme.

The DFT computations were performed using the code Wien2K,\cite{PBLA99} which is based on a full-potential linearized
augmented plane wave (FLAPW) method. Pd, Ni and all the solid
solutions crystallize in an fcc lattice with space group
Fm$\bar{3}$m.\cite{dd} $x$ = 0, 0.25, 0.5, 0.75 and 1 were taken
for the computations. Pd and Ni ($x$ = 0, 1) structures were first
constructed by taking literature values of the respective lattice
constants, and then by optimizing the volume, so that the
equilibrium lattice constants correspond to the respective minimum
energy configurations. Birch-Murnaghan equation of state
\cite{birch,murnaghan} was used to fit the energy versus volume
curves for the optimizations. The crystal structure for $x$ = 0.25
(0.75) was generated by constructing a $1\times1\times2$ supercell
of Pd (Ni) and then replacing one of the four site-split Pd (Ni)
atoms with Ni (Pd). For the case of $x$ = 0.5, 2 of the four
site-split Pd atoms in a $1\times1\times1$ supercell were replaced
with Ni. All these structures were separately volume-optimized by
the same procedure as stated above. For the $x \neq 0$ structures,
the atomic coordinates were further relaxed to limit the atomic
forces to less than 1 mRy/au. The exchange-correlation functionals
adopted for the calculations were taken according to the
generalized gradient approximation (GGA) as introduced by Perdew,
Burke and Ernzerhof.\cite{perdew} The energy of separation
between core and valence states was taken as -0.6 Ry. In the FLAPW
method, the potential in a Muffin-tin radius R$_{MT}$ around each
atom is taken as atomic-like, and atomic spherical wavefunctions
are used as basis functions for the basis set,\cite{cottenier}
while in the interstitials, the potential is smooth and plane
waves constitute the basis functions. The R$_{MT}$ value for both
Pd and Ni was taken as 2.5 a.u. In all the calculations, the
wavevectors for plane waves were kept limited to a maximum value
k$_{max}$ such that k$_{max}$R$_{MT}$ = 7.0. The maximum
multipolarity \emph{l}$_{max}$ for the spherical wavefunctions was
set at 10. Further, the Fourier expansion of the charge density
was limited to G$_{max}$ = 12. The Brillouin zone was sampled
using a k-mesh with 72 irreducible k-points.

From the eDOS's as computed from the DFT method described above,
$C_e(T_e)$ and $G(T_e)$ were calculated using equations (6) and (7)
for each composition. For the latter, in addition,
$\lambda$$\left\langle \omega^2\right\rangle$ values were taken
from literature for pure Pd and Ni, which were then linearly
interpolated to get the values for intermediate compositions. For
the TTM calculations in the next step, $C_e(T_e)$ and $G(T_e)$
values for temperature ranges taken usually in a TSM code
\cite{ff} were sampled out from the derived $C_e(T_e)$ and
$G(T_e)$ curves, and used as inputs to the code. The rest of the
parameters, viz. $K_e$, $C_l$ and $K_l$ were interpolated between
Pd and Ni values in the same manner as $\lambda$$\left\langle
\omega^2\right\rangle$ values were derived for calculating $G(T_e)$.

%\section{Experimental}

\section{Results and discussion}
\subsection{XRD}

X-ray diffraction patterns of all the samples before and after
irradiation are shown in Figs. 1 (a) - (d). Peaks corresponding to
Pd, Ni and Si elemental solids have been identified from the
corresponding Joint Committee on Powder Diffraction Standards
(JCPDS) data. The peaks of the compounds, on the other hand, have
been identified by a comparison with the powder diffraction
patterns simulated using PowderCell\cite{PowCel} and by taking
the crystal parameters for NiSi, Ni$_3$Si$_2$ and Ni$_2$Si from Ref. [30] and for PdSi\cite{PdSiXSt} and Pd$_2$Si\cite{Pd2SiXSt} from SpringerMaterials online database. A look at the XRD patterns of the pristine samples indicates the presence of a combination of elemental and compound phases mentioned above. The elemental peaks are as expected, while the compounds (silicides) must have been formed during the film deposition. Silicide formation while depositing thin metal films on Si is not uncommon.\cite{gg}

\begin{figure}%
\centering
\subfigure[][]{%
\label{fig:2-a}%
\includegraphics[width=0.42\textwidth, trim = 60 10 0 60, clip = true]{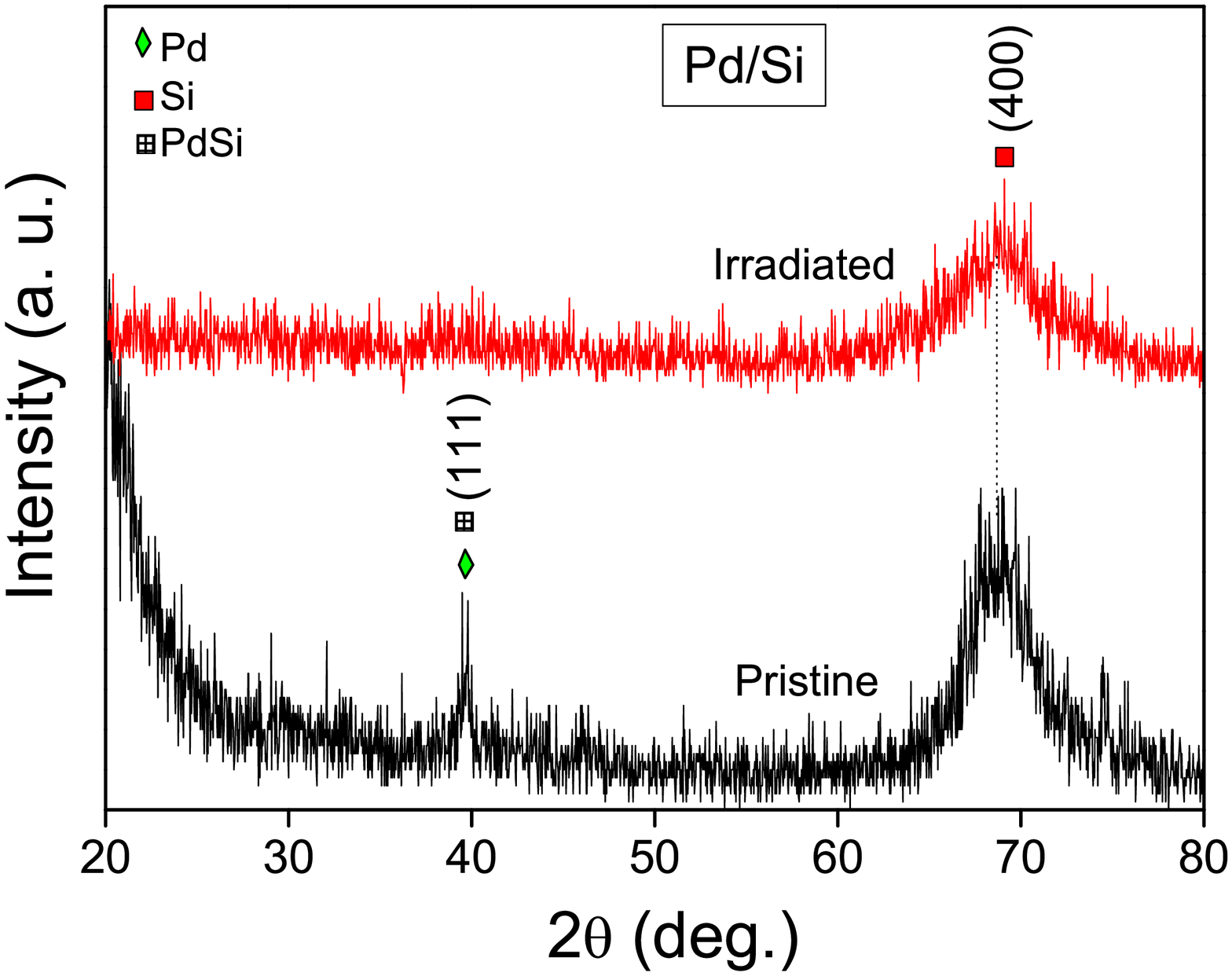}}%
\hspace{5pt}%
\subfigure[][]{%
\label{fig:2-b}%
\includegraphics[width=0.42\textwidth, trim = 60 10 0 60, clip = true]{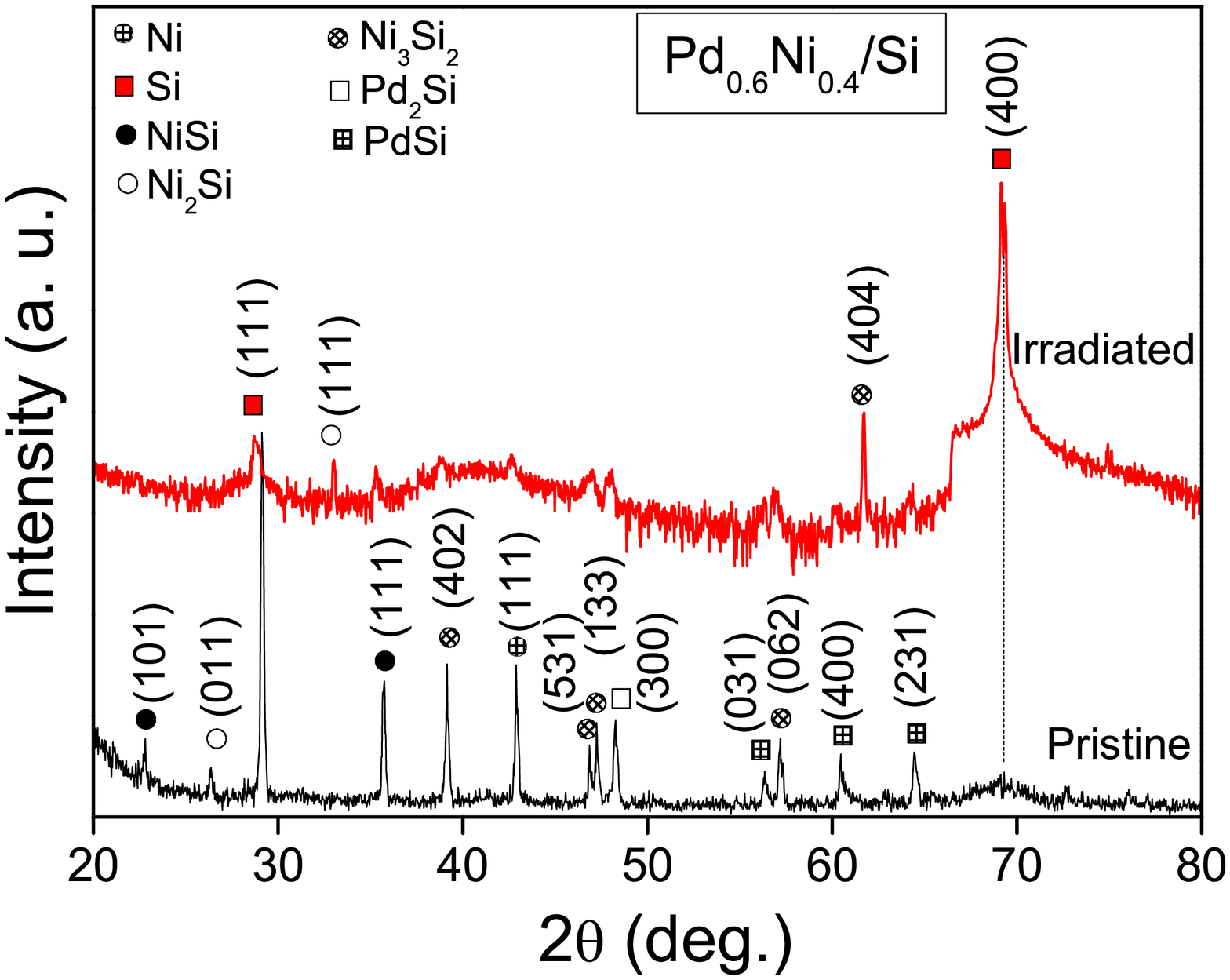}}%
\hspace{5pt}%
\subfigure[][]{%
\label{fig:2-c}%
\includegraphics[width=0.42\textwidth, trim = 60 10 0 60, clip = true]{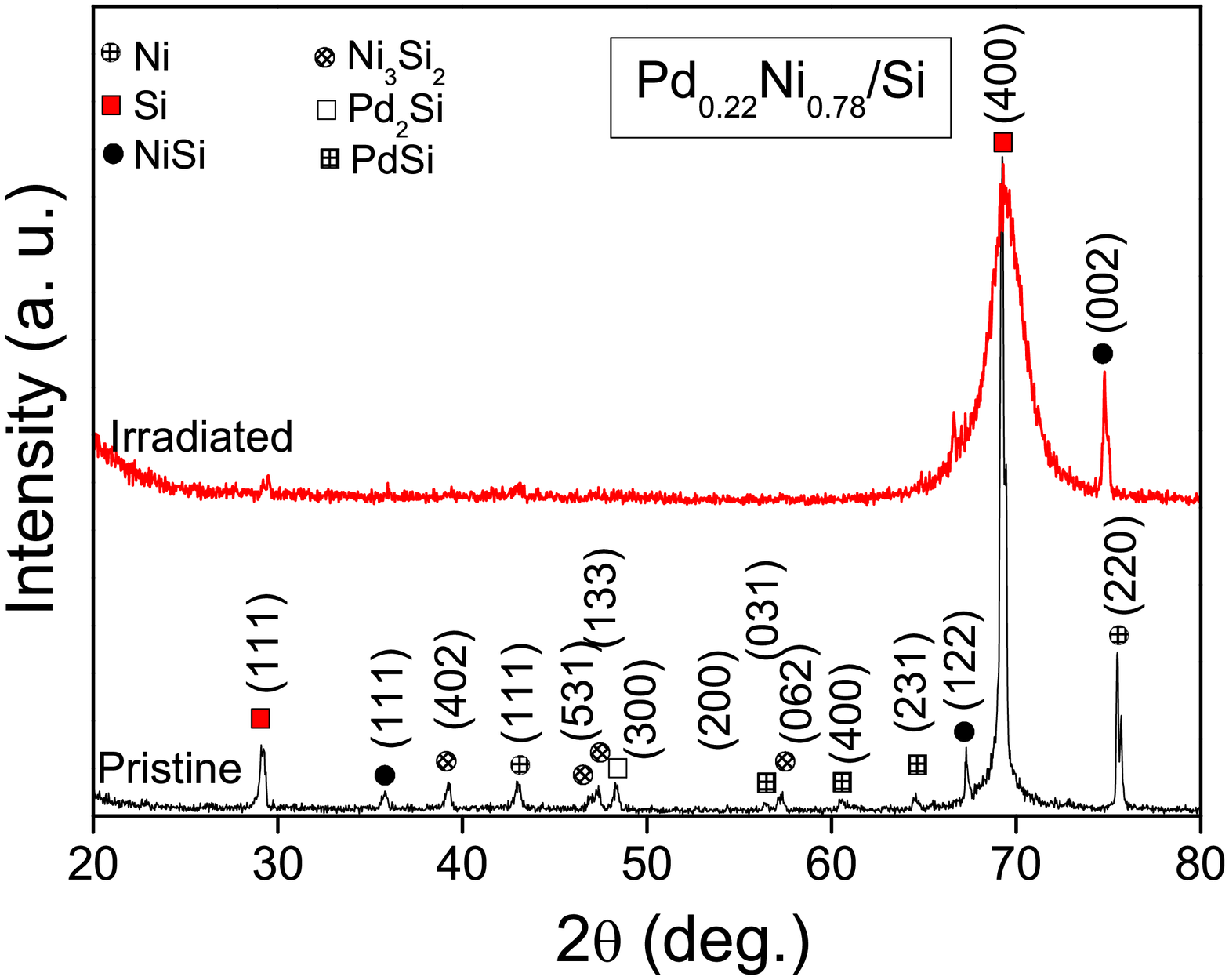}}%
\hspace{5pt}%
\subfigure[][]{%
\label{fig:2-d}%
\includegraphics[width=0.42\textwidth, trim = 60 10 0 60, clip = true]{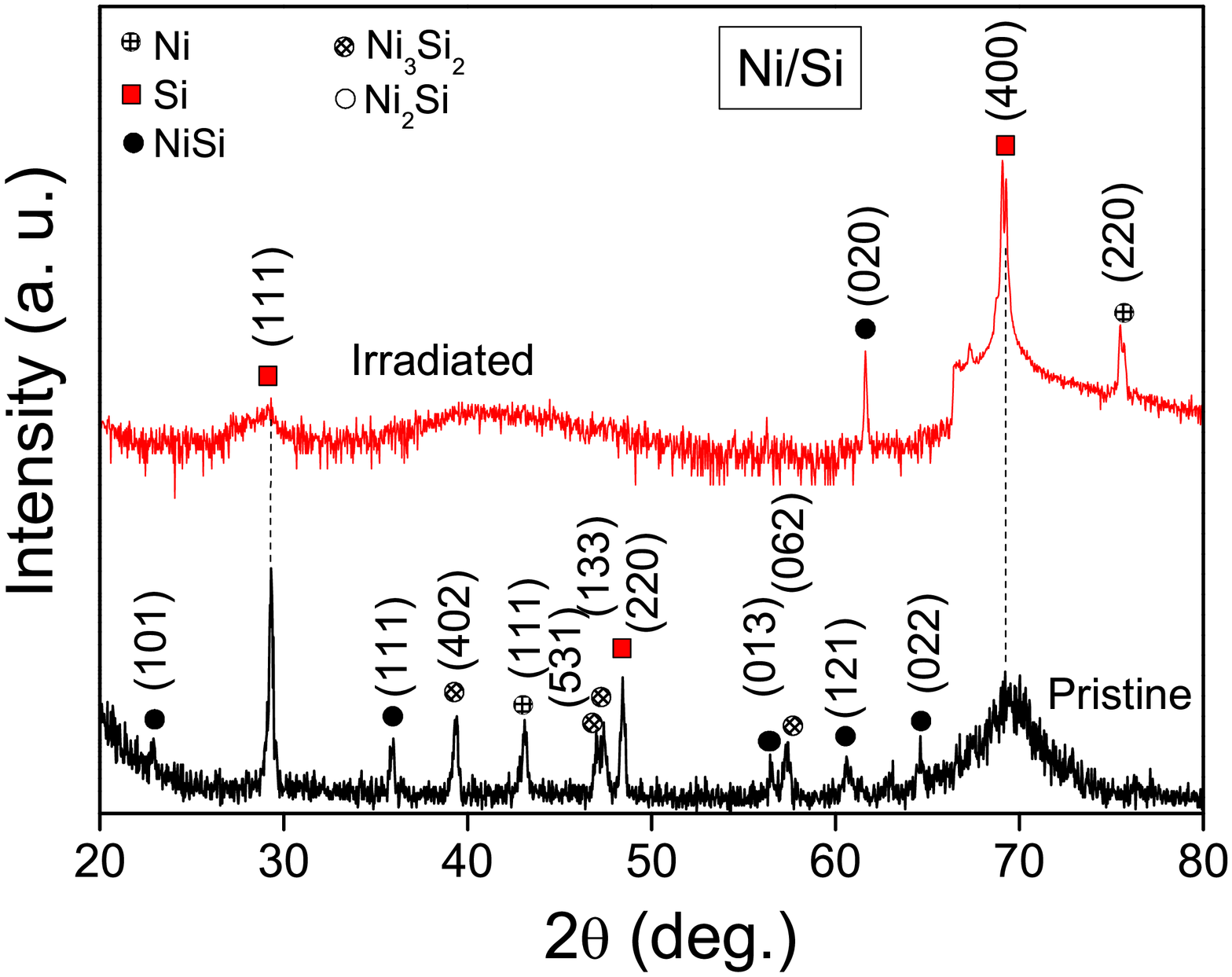}}%

\caption[]{XRD spectra of (a) Pd/Si (b) Pd$_{0.22}$Ni$_{0.78}$/Si
(c) Pd$_{0.60}$Ni$_{0.40}$/Si and (d) Ni/Si. The steps at $\sim$
67$^{\circ}$ in (b) and (d) are experimental artefacts, and do not
affect the interpretations made.}

\label{fig:5}%
\end{figure}

The common effect of irradiation of all the samples has been
either a complete removal or a significant suppression or
broadening of almost all but the substrate peaks. This must be due
to SHI irradiation induced defect creation and amorphization of
the films. Such irradiation induced effects are also not uncommon.\cite{hh} The takeaway message from the XRD patterns is that (i)
the samples contain thin films of a combination of elemental
and silicide phases, and (ii) SHI irradiation does produce
structural modifications in the samples. Any further
interpretation of the XRD patterns would perhaps become an
over-interpretation.

\subsection{RBS}

RBS spectra of all the pristine and irradiated samples were
recorded to determine the compositions and thicknesses of the thin
films, and also to examine whether there is any interface mixing
occurring as a consequence of the irradiation alone. The spectra
and their fits using the code SIMNRA\cite{SIMNRA} are shown in
Figs. 2(a) - (d). According to the fits, the samples have
configurations Pd (19.2 nm)/Si, Pd$_{0.60}$Ni$_{0.40}$ (28.8
nm)/Si, Pd$_{0.22}$Ni$_{0.78}$ (18.8 nm)/Si and Ni (23.0 nm)/Si.
The spectra were fitted using a resolution in the range of 20 - 24
keV, which is equivalent to $\sim$ 12 nm. An insignificant but
noticeable decrease in Pd and Ni peak intensities after
irradiation for all the samples indicates that Pd/Si and Ni/Si
interface mixing might have taken place as a consequence of
irradiation. However, the mixing thicknesses must be too small
compared to the resolution ($\sim$ 12 nm) to show up in the
spectra to any significant extent. In order to investigate whether
there is indeed a SHI induced mixing, XPS depth profiles, which
could provide a much better spatial resolution, have been
performed on all the pristine and irradiated samples. The results
are discussed in the following section.

\begin{figure}%
\centering
\subfigure[][]{%
\label{fig:2-a}%
\includegraphics[width=0.4\textwidth]{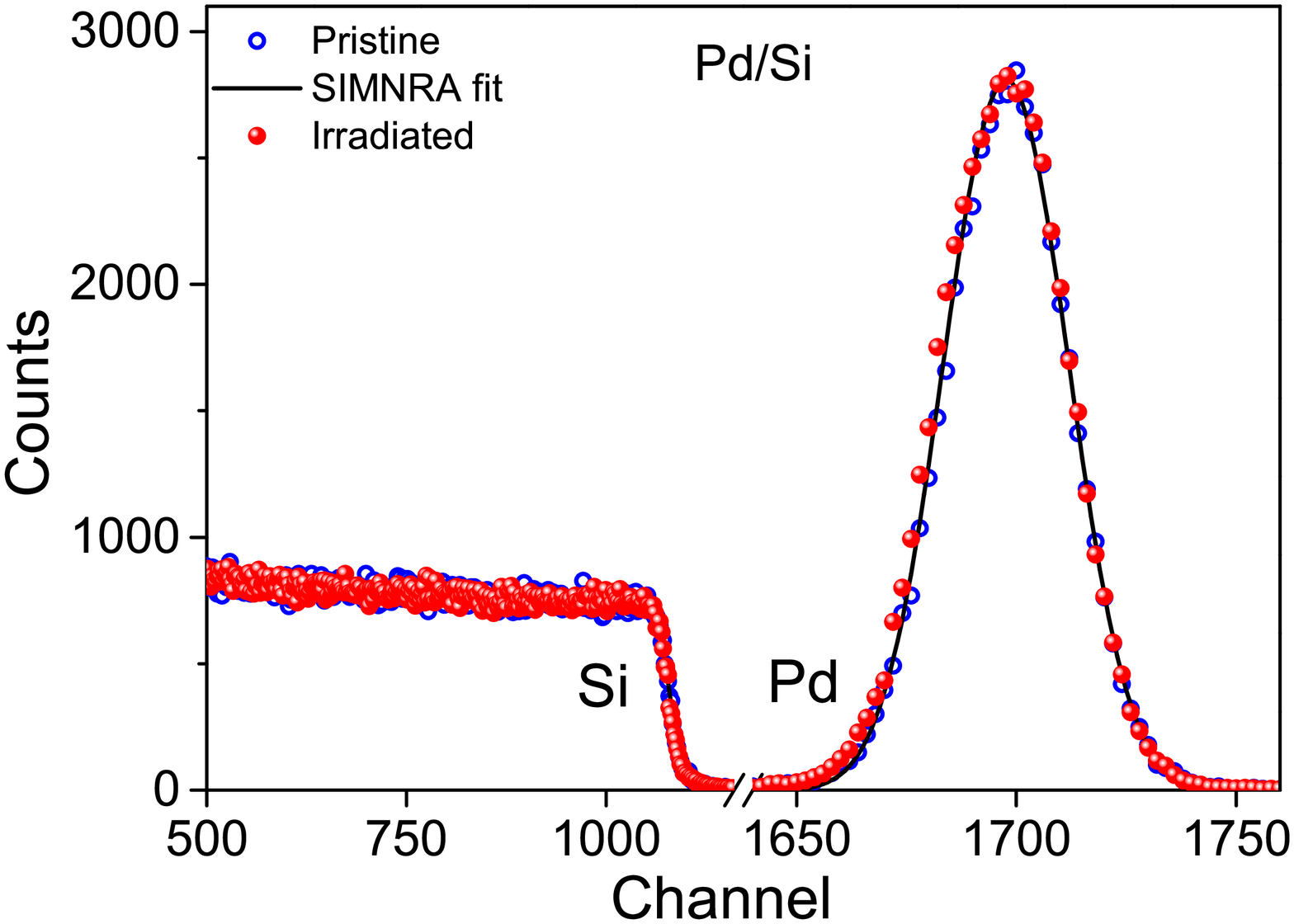}}%
\hspace{5pt}%
\subfigure[][]{%
\label{fig:2-b}%
\includegraphics[width=0.4\textwidth]{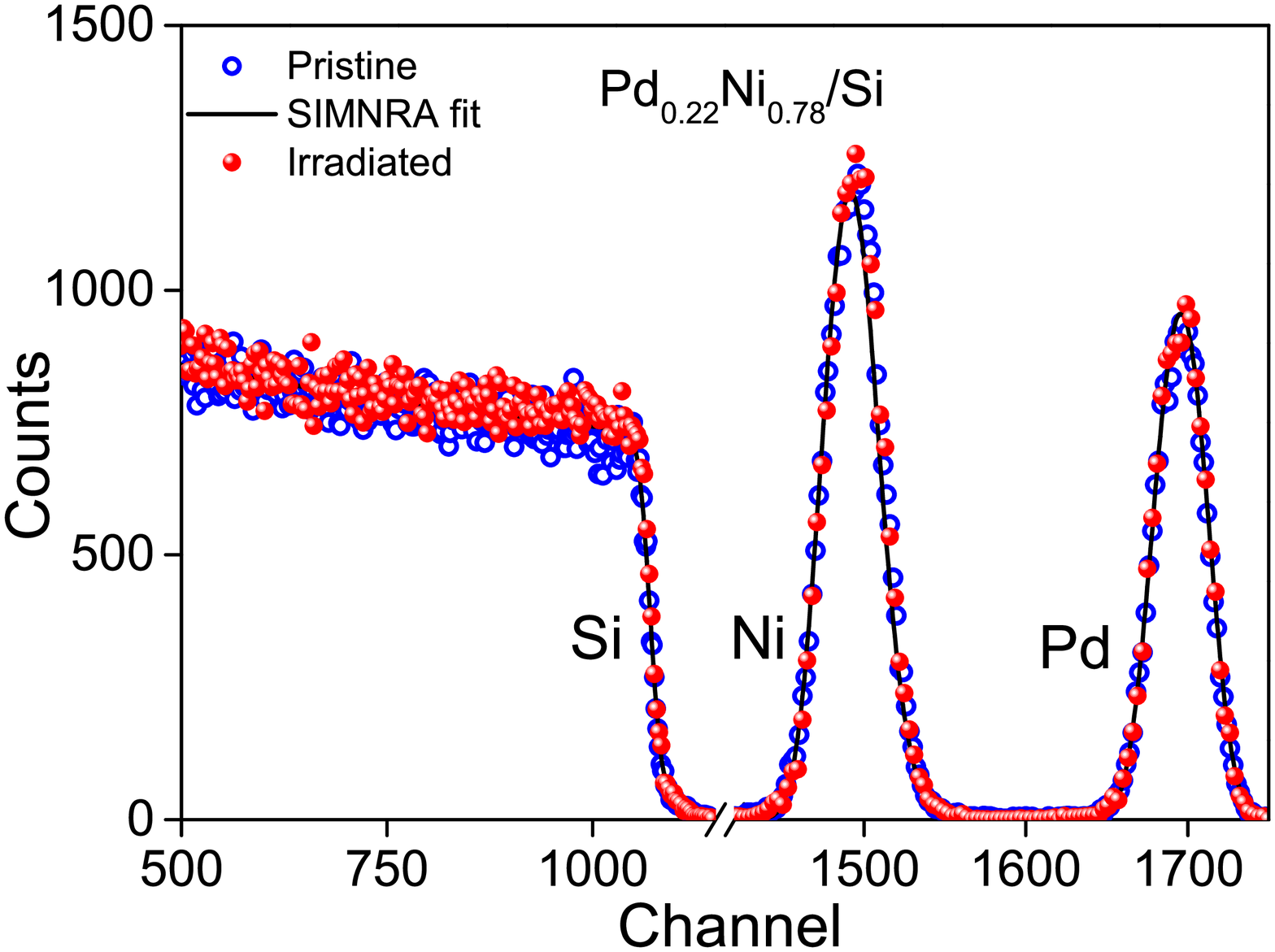}}%
\hspace{5pt}%
\subfigure[][]{%
\label{fig:2-c}%
\includegraphics[width=0.4\textwidth]{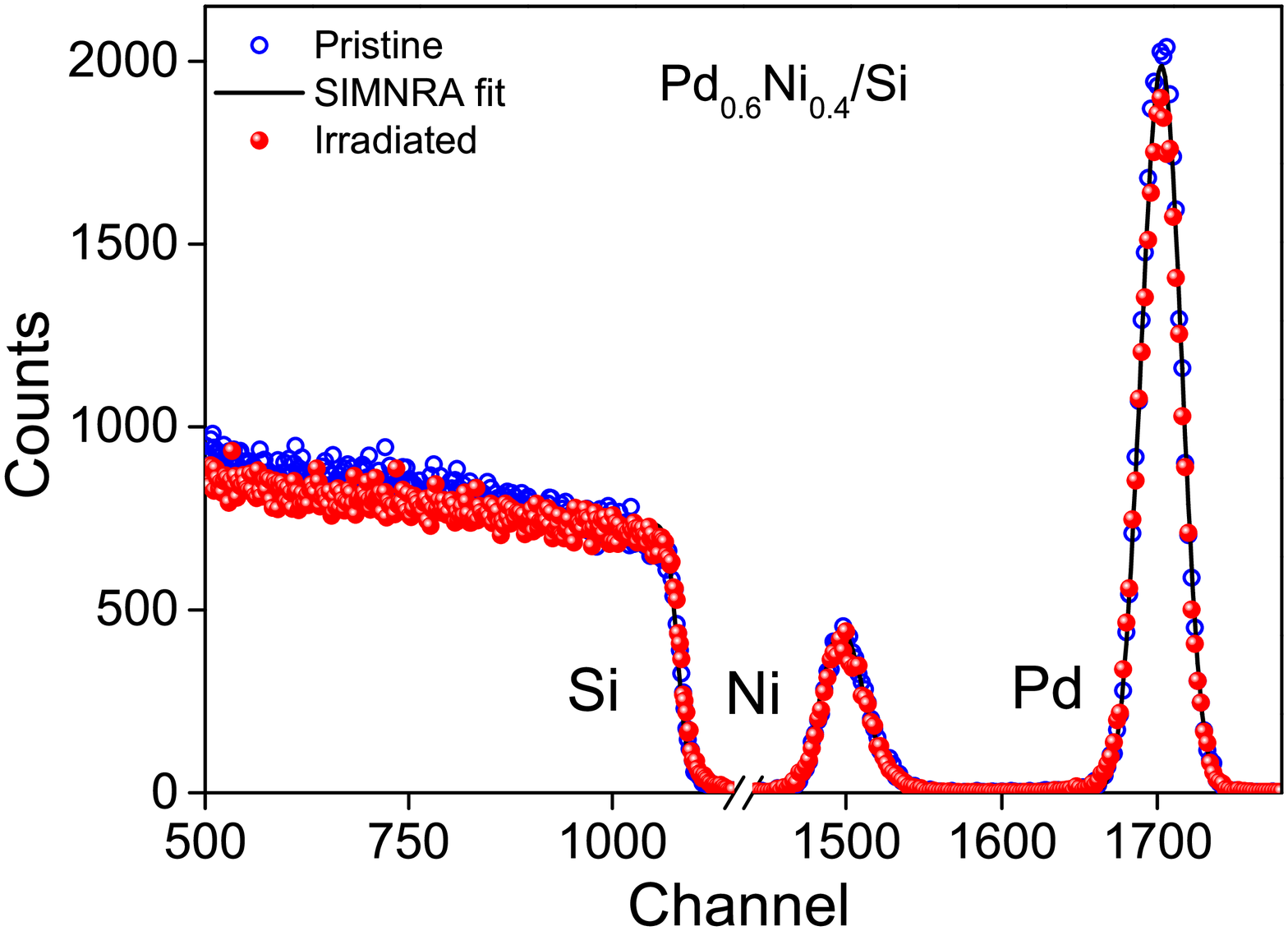}}%
\hspace{5pt}%
\subfigure[][]{%
\label{fig:2-d}%
\includegraphics[width=0.4\textwidth]{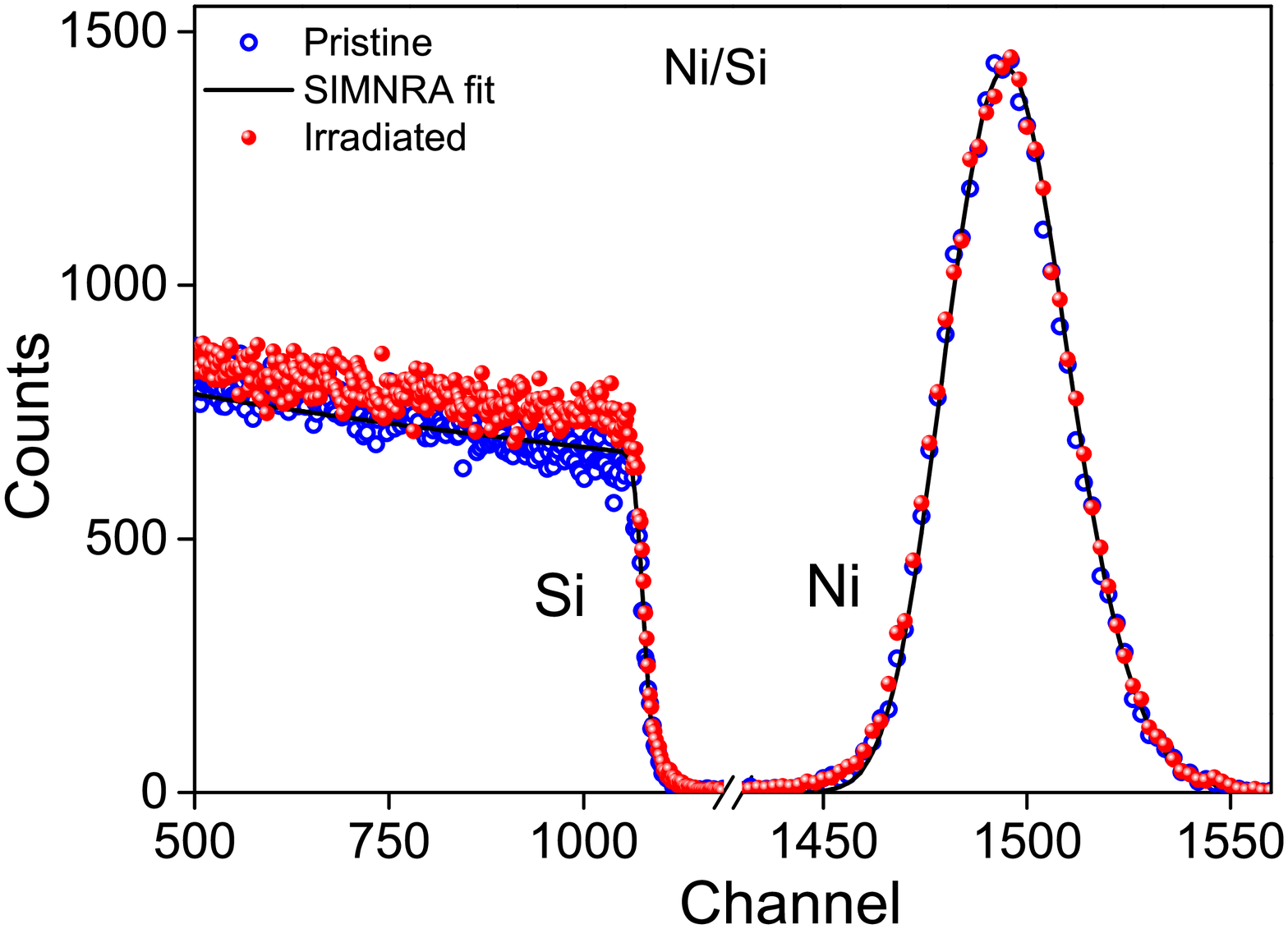}}%

\caption[]{RBS spectra of (a) Pd/Si (b) Pd$_{0.22}$Ni$_{0.78}$/Si
(c) Pd$_{0.60}$Ni$_{0.40}$/Si and (d) Ni/Si}
\label{fig:5}%
\end{figure}

\subsection{XPS depth profile}

All the samples have been analyzed using depth profiling XPS
paired with 1 keV Ar$^+$ ion sputtering to collect Pd3d, Ni2p and
Si2p high-resolution spectra. A large number of sputtering cycles
were used for the study so that the film/substrate interface is
reached in about 35 - 40 cycles. A comparison of these many number
of sputter cycles (to reach the interface) with the depth of the
interface (i.e., the film thickness) indicates that the XPS
spectra have been recorded at the interval of 0.5 - 0.6 nm. This
is likely to provide a (minimum) depth resolution of 0.5 - 0.6 nm
for elemental depth profiling, a spatial depth resolution at least
an order of magnitude better than the RBS depth resolution
mentioned above, and hence to enable us to observe any small
amount of SHI induced interdiffusion (or intermixing) of Pd or Ni
into Si and vice-versa. It should be noted here, though, that the
Ar$^+$ ion sputtering process itself can induce an additional
interface mixing.\cite{Hof98} However, since its effects are
similar for all the samples, any $x$-variation of observed mixing
can be taken essentially as the SHI irradiation effect, which is
augmented further by the sputtering induced mixing equally. We
will be comparing the experimentally observed $x$-variation of
mixing with the $x$-variation of an equivalent quantity estimated
using DFT and TTM computations.

Figure 3 shows the high-resolution XPS spectra in Pd 3d$_{5/2}$,
Ni 2p$_{3/2}$ and Si 2p$_{3/2}$ regions of the pristine and
irradiated Pd$_{0.60}$Ni$_{0.40}$/Si samples for different sputter
cycles. This composition has been taken as a representative for
all the samples. For both the pristine and irradiated samples,
both the Pd and Ni XPS peaks (i) diminish gradually and (ii) shift
to higher binding energies (BE's), while approaching the interface.
The Si peak also has a gradual rise from the interface, with a
less pronounced shift. The gradual changes in the peak heights are
indicative of interdiffusion or intermixing of Pd and Ni in Si
even in the pristine sample. The intermixed interface seems to
broaden on irradiation, as can be seen from the apparently deeper
interpenetrations of the peaks in the interfacial region. These
observations can be made more quantitative by plotting the XPS
peak position at maximum intensity (PPMI) versus sputter cycle
(SC).

\begin{figure}%
\centering
\subfigure[][]{%
\label{fig:3-a}%
\includegraphics[width=0.58\textwidth, trim = 30 180 0 20, clip = true]{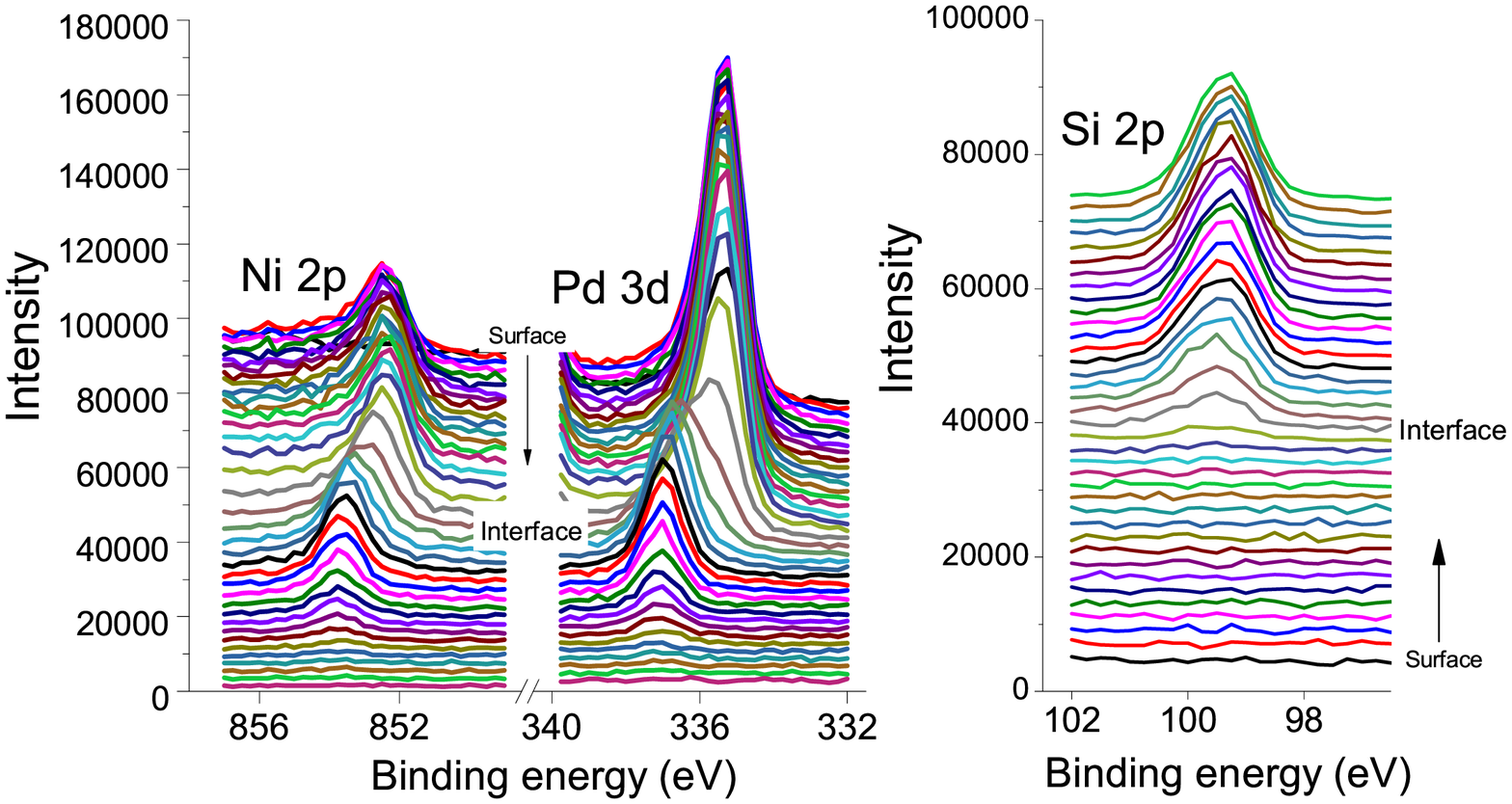}}%
\hspace{5pt}%
\subfigure[][]{%
\label{fig:3-b}%
\includegraphics[width=0.58\textwidth, trim = 30 180 0 20, clip = true]{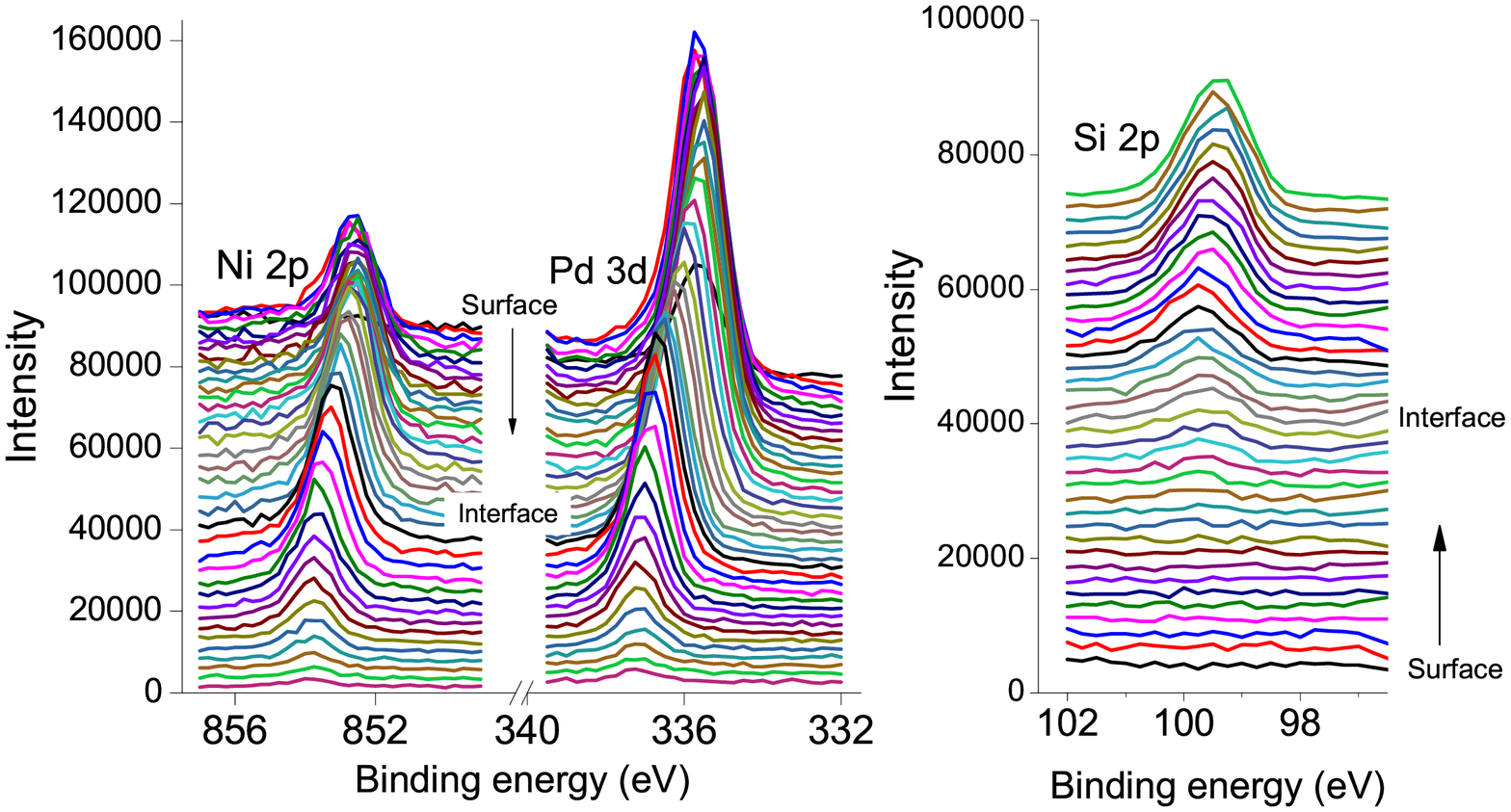}}%
\hspace{5pt}%

\caption[]{High-resolution XPS spectra in Pd 3d$_{5/2}$, Ni
2p$_{3/2}$ and Si 2p$_{3/2}$ regions of Pd$_{0.60}$Ni$_{0.40}$/Si
pristine (a) and irradiated (b) samples for different sputter
cycles. The curves corresponding to the surface and interface are
indicated. For clarity, the spectra of alternate cycles only have
been plotted.}
\label{fig:3}%
\end{figure}

Figure 4 displays the PPMI versus SC plots in Pd 3d$_{5/2}$, Ni
2p$_{3/2}$ and Si 2p$_{3/2}$ regions of the pristine and
irradiated Pd$_{0.60}$Ni$_{0.40}$/Si samples. The symbol sizes are
proportional to the corresponding normalized intensities. It is to
be noted that the PPMI's as plotted in the figure are not true
single peak positions, as each of the Pd 3d$_{5/2}$, Ni 2p$_{3/2}$
and Si 2p$_{3/2}$ peaks may consist of more than one sub-peaks
signifying different elemental or compound phases. A deconvolution
of these peaks, which will be shown in the following, would
identify the phases present. Coming back to the PPMI versus SC
plots, the PPMI for pristine Ni shifts continuously from 852.25 eV
at SC 32 to 853.75 at SC 47 with concomitantly diminishing
intensity. The trend continues beyond the 47$^{\rm {th}}$ cycle
with the PPMI saturating at 854 eV. The concomitant PPMI shift and
intensity reduction is indicative of an increase in the number of
Si atoms surrounding a Ni atom, and hence suggests the occurrence
of diffusion of Ni into Si substrate.\cite{Kumar81,Kumar16} The
PPMI profile with SC, thus, can be considered to represent the
reverse Ni concentration profile with depth (depth profile). The
middle (around SC 40) of the interfacial interdiffused region (SC
32 to SC 47) can be considered as the true film substrate
interface. Equating SC 40 to 28.8 nm as obtained from the RBS
results, the interdiffsuion region extends from about 23.0 nm to
about 33.8 nm, i.e. in a span of about 10.8 nm. This can be
considered as the standard deviation $\sigma$ of the interfacial
position, or in other words the interface width, for the present
depth profile. This Ni-Si interdiffusion has taken place during
the deposition itself, as has also been argued in the XRD
subsection above. The effect of 100 MeV Au irradiation, augmented
by the Ar$^+$ ion sputtering during the depth profiling, has been
to broaden the PPMI profile such that the PPMI shifts from 852.25
eV at SC 30 to 853.75 at SC around 55, once again with
continuously diminishing intensity. The standard deviation after
irradiation, thus, becomes 25 sputter cycles, which is equivalent
to 18.0 nm. As the irradiation itself has been shown to cause
modifications in the sample as revealed from the XRD patterns, the
irradiation can be considered as the primary source of the
relative interface broadening or the enhanced interdiffusion. The
enhanced interdiffusion, in turn, can be considered to represent
the 100 MeV Au (i. e., SHI) induced mixing. The PPMI profile of Pd
also follows the same pattern, the only difference being that the
$\sigma$ value changes from 11 cycles (9.8 nm) for the pristine
sample to 26 cycles (19.4 nm) after irradiation. In the case of
Si, the intensity starts decreasing from about the same depth till
which Pd and Ni interdiffuse for both pristine and irradiated
samples separately. Further, the presence of Si extends, with
diminishing intensity, into the film till the depth from where Pd
and Ni had started depleting. Although the Si peak shifts are
small, and hence likening it to a concentration versus depth
profile would not be very convincing, its simultaneous presence
with Pd and Ni corroborates the conjecture of interdiffusion of Pd
and Ni in Si, which increases as a result of irradiation. The
interfacial broadening can be better quantified by determining the
atomic fractions of the elements using the areas under the peaks
and the elemental sensitivity factors, and then plotting these
against the depth, scaled appropriately from the sputter cycle.
These depth profiles for all the pristine and irradiated samples
will be shown and discussed later.

\begin{figure}%
\centering
\includegraphics[width=0.4\textwidth]{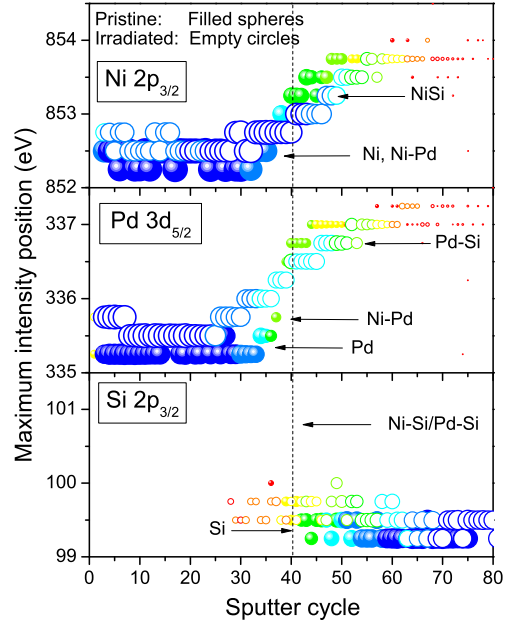}%
\hspace{5pt}%

\caption[]{Binding energy positions of maximum intensity versus
sputter cycle in Pd 3d$_{5/2}$, Ni 2p$_{3/2}$ and Si 2p$_{3/2}$
regions of Pd$_{0.60}$Ni$_{0.40}$/Si pristine and irradiated
samples. The symbol sizes and colours signify the intensity of the
maximum for a particular sputter cycle normalized with respect to
the maximum intensity which is maximum of all for each region and
sample condition (pristine or irradiated). Tentative elements or
compound phases corresponding to different peak positions are also
shown for a guidance. The apparent film substrate interface is
marked with a dashed line.}
\label{fig:4}%

\end{figure}

%%%%%%%%

It would be worthwhile in the meantime to examine the XPS spectra
of these samples at the interface (SC 40) to see the interfacial
phases present before and after irradiation. The background
corrected high-resolution XPS spectra in Si 2p$_{3/2}$, Pd
3d$_{5/2}$, O 1s and Ni 2p$_{3/2}$ regions and their fits for the
Pd$_{0.60}$Ni$_{0.40}$/Si pristine and irradiated samples are
shown in Fig. 5. Tentative assignments of different peak positions
to pure\cite{NIST} Pd and Ni and their silicides with random
compositions (e.g., Pd$_x$Si$_{1-x}$\cite{YTakagi,QWei,YCao}) or
in compound form, like Ni$_2$Si\cite{YCao,XChen}, and to adsorbed
oxygen,\cite{HGabasch} are also shown. In brief, both the
pristine and irradiated Pd$_{0.60}$Ni$_{0.40}$/Si samples contain
elemental and silicide phases of Pd and Ni at the interface with
slightly different amounts before and after irradiation. The
irradiated sample has an additional Pd$_x$Si$_{1-x}$ peak. These
conjectures are in agreement with the XRD observations, as
discussed earlier.

%%%%%
\begin{figure}%
\centering
\includegraphics[width=0.8\textwidth, trim = 60 20 40 40, clip = true]{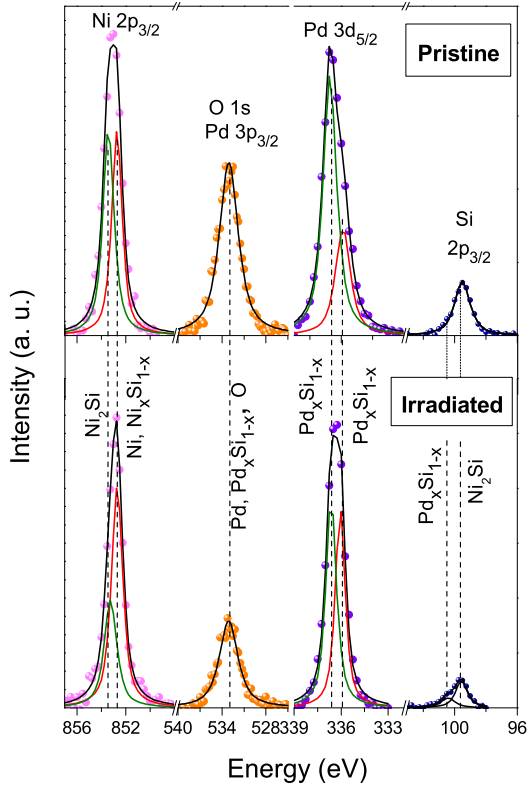}%
\hspace{5pt}%

\caption[]{The background corrected high-resolution XPS spectra in
Si 2p$_{3/2}$, Pd 3d$_{5/2}$, O 1s and Ni 2p$_{3/2}$ regions
(shown in top panel) and their fits for the
Pd$_{0.60}$Ni$_{0.40}$/Si pristine and irradiated samples. O1s and
Pd 3p$_{3/2}$ spectra overlap \cite{HGabasch}. Tentative
assignments of different peak positions (bottom panel) are also
shown.}
\label{fig:5}%

\end{figure}

%%%%%%

Moving on to the depth profiles, these are shown in Figs. 6 (a) -
(d) as variations of atomic fractions (concentrations, $m$) of Pd,
Ni and Si as a function of depth $z$ for the four studied samples.
Here, the SC has been converted to depth with the help of RBS
analyses as discussed above. The depth profiles have been obtained
using the quantification scheme provided in the MultiPak Data
Reduction Software available with the PHI 5000 Versaprobe II XPS
instrument used for measuring XPS spectra. Absence of sharp
interfaces even in the pristine samples, as can be seen from
figures, suggests that there is already an interdiffusion in the
pristine samples, in line with the earlier arguments. This
interdiffusion might have taken place during the film deposition,
and may have further been augmented by Ar$^+$ ion sputtering
during the depth profiling.

The Pd and Ni depth profiles have been roughly fitted with error
function, and the fits are shown overlapping with the data. The
squared interface width, or variance, ${\sigma}^2$ has then been
calculated using \cite{Munster2004}

\begin{equation}
 \sigma^2 = {{\int_{0}^{\infty}z^2 m' (z) dz} \over {\int_{0}^{\infty} m' (z) dz}},
\end{equation}
where $m' (z)$ is the gradient of the fitted interface profile.
The change $\Delta \sigma^2$ in variance on irradiation for a
particular depth profile is then given by the difference in the
variances after and before irradiation. The value 265$\pm$30 nm of
$\Delta \sigma^2$ for Ni in Si for the case of $x$ = 0.40 can be
compared with (18 nm)$^2$ - (10.8 nm)$^2$ $\sim$ 208 nm$^2$, as
obtained using the PPMI versus SC plots discussed above. These
values are fairly close to each other.

\begin{figure}%
\centering
\subfigure[][]{%
\label{fig:5-a}%
\includegraphics[width=0.35\textwidth, trim = 40 00 40 40, clip = true]{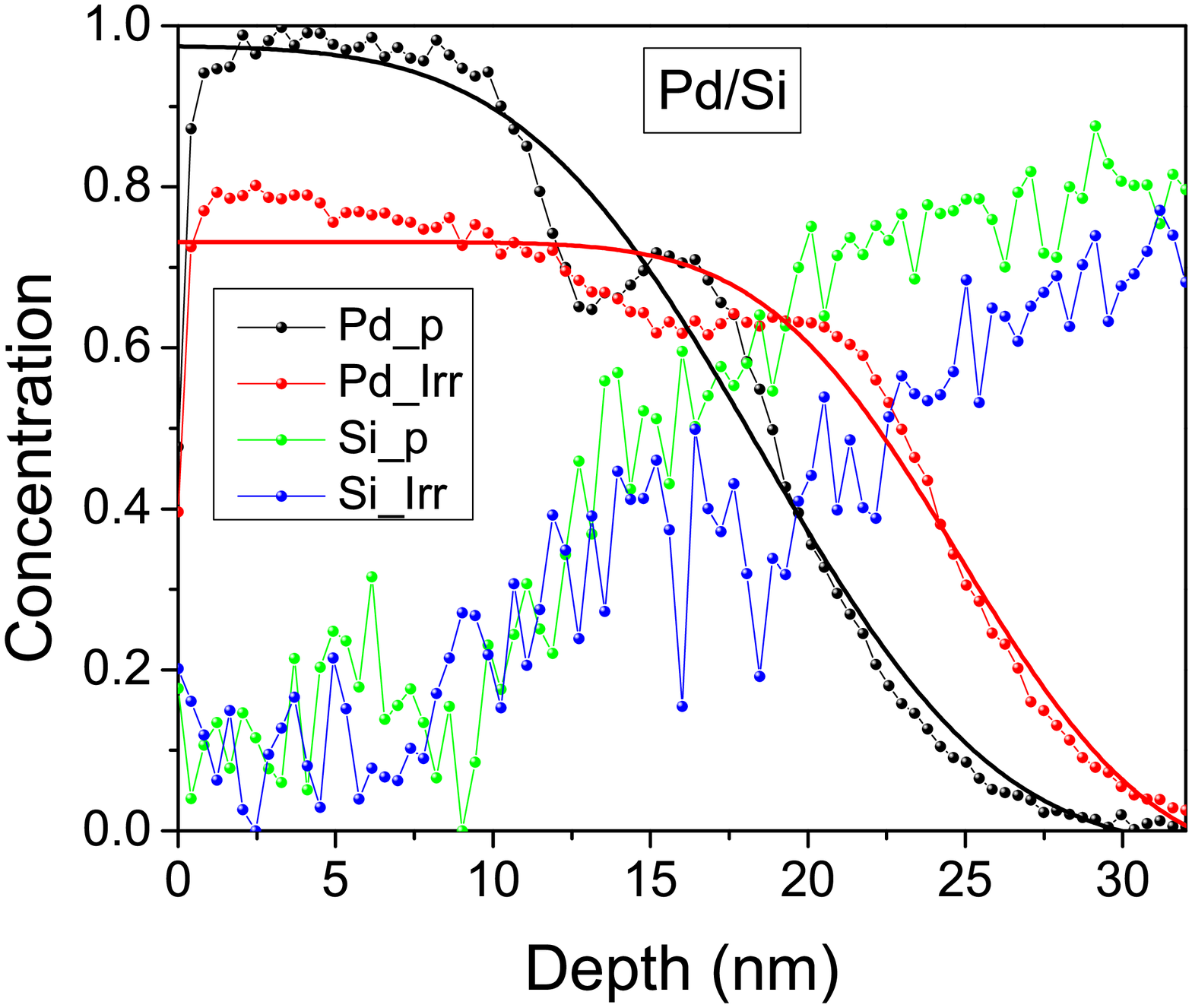}}%
\hspace{8pt}%
\subfigure[][]{%
\label{fig:5-b}%
\includegraphics[width=0.35\textwidth, trim = 40 00 40 40, clip = true]{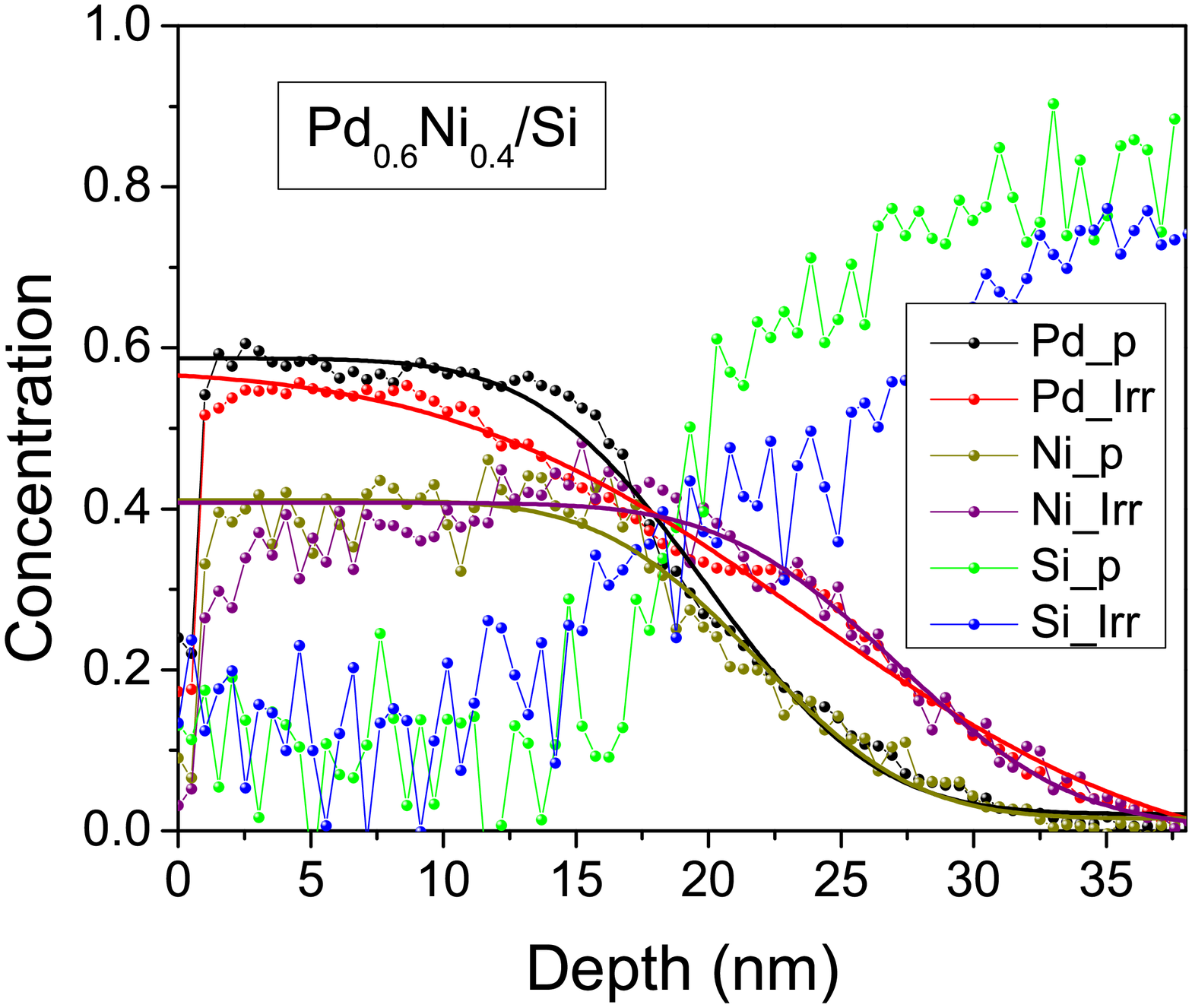}}%
\hspace{8pt}%
\subfigure[][]{%
\label{fig:5-c}%
\includegraphics[width=0.35\textwidth, trim = 40 00 40 40, clip = true]{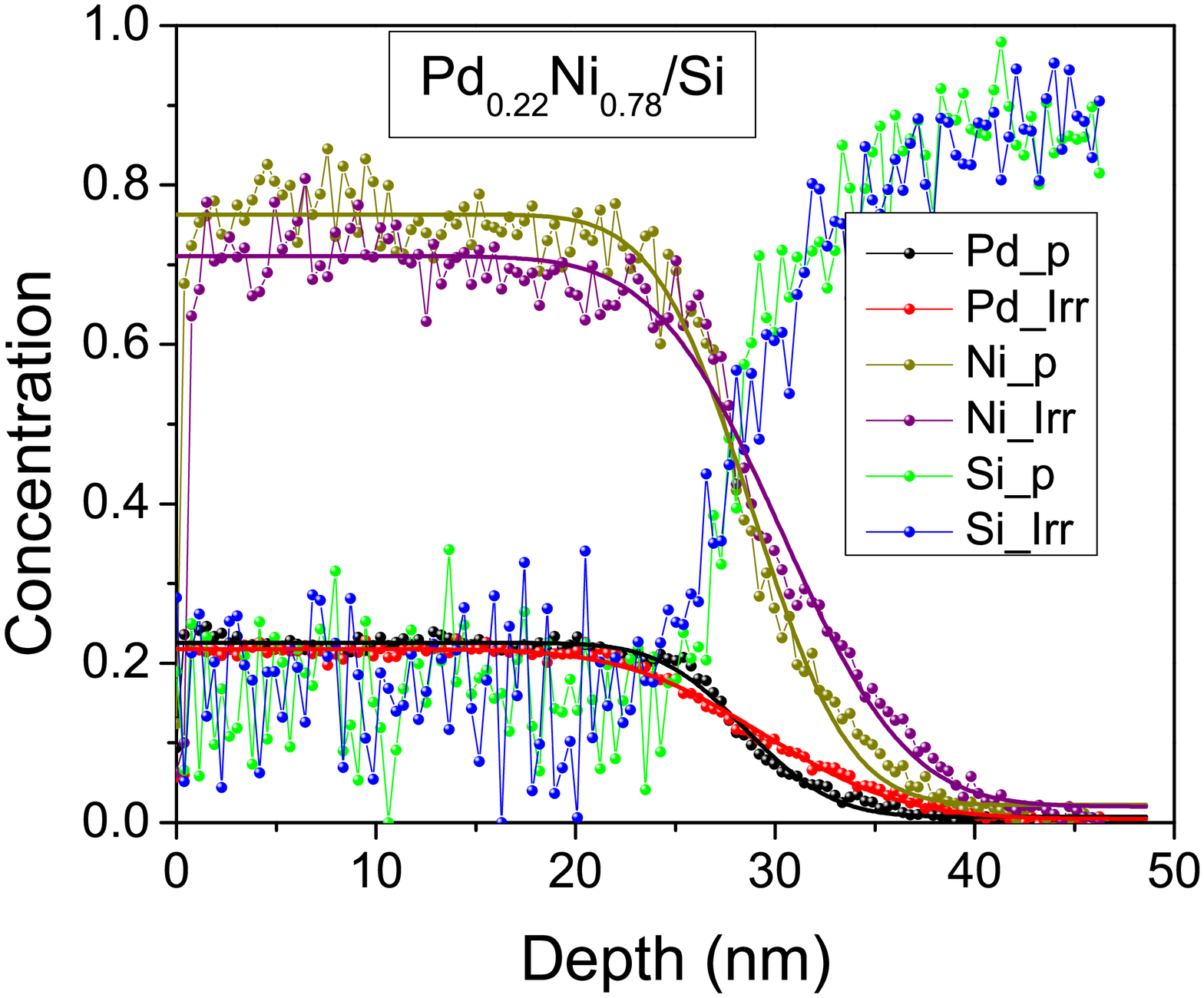}}%
\hspace{8pt}%
\subfigure[][]{%
\label{fig:5-d}%
\includegraphics[width=0.35\textwidth, trim = 40 00 40 40, clip = true]{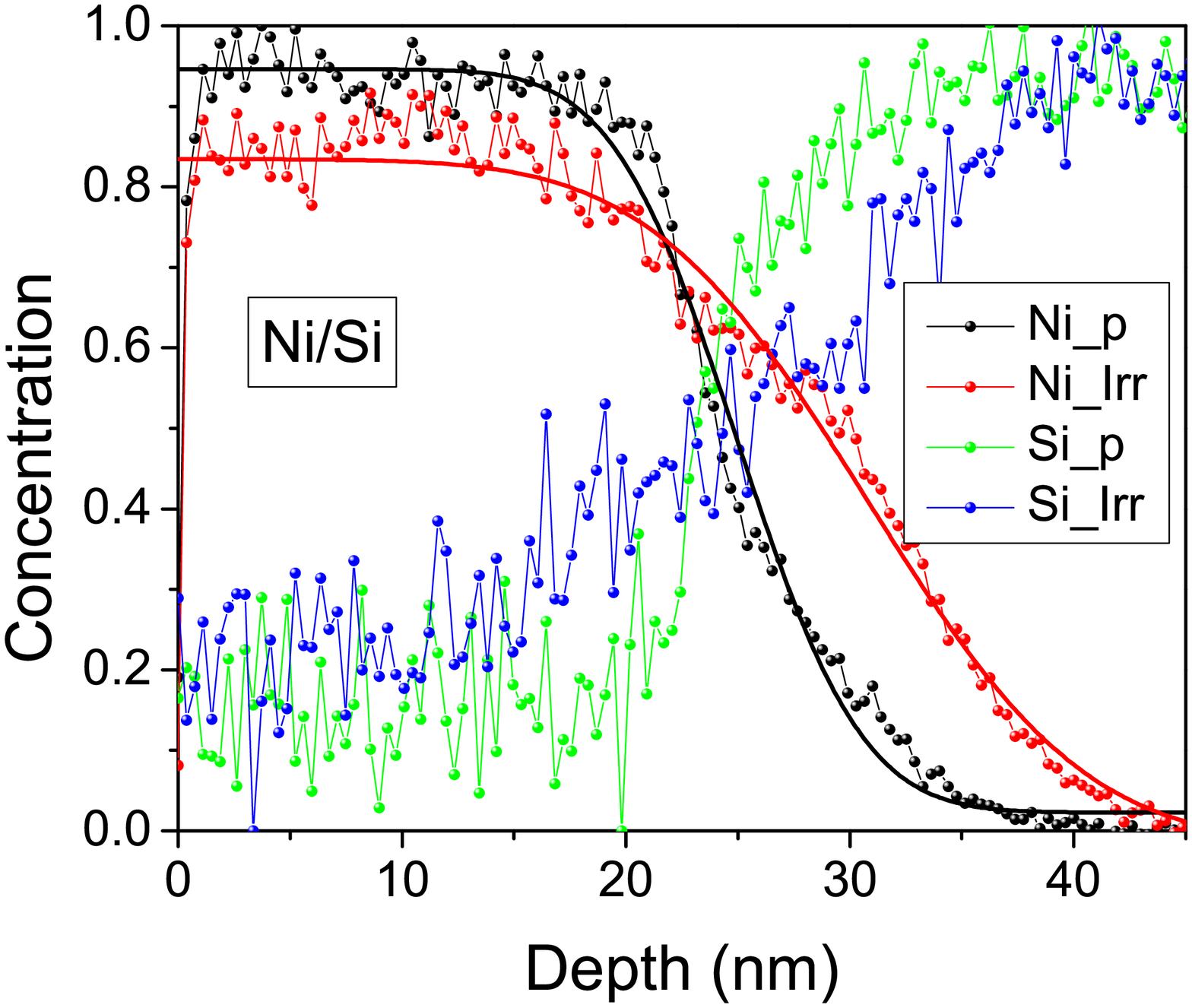}}%

\caption{XPS depth profile of (a) Pd/Si (b)
Pd$_{0.22}$Ni$_{0.78}$/Si (c) Pd$_{0.60}$Ni$_{0.40}$/Si and (d)
Ni/Si. Overlapped with Pd and Ni profiles are their error function
fits.} \label{fig:four graphs}
\end{figure}

Figure 7 shows the variation of $\Delta \sigma^2$, for Pd and Ni
interdiffusion in Si, with Ni concentration $x$. It is noteworthy
that $\Delta \sigma^2 (x)$ is neither flat nor monotonic; it
rather has a pronounced minimum. The crudest approximation to this
variation would be a curve which is concave upward and has a
rather deep minimum. In the following sections, we would attempt
to examine whether such a variation can be simulated
computationally based on the concepts of TSM and DFT.

\begin{figure}[t]
\vspace{0.0cm}
\includegraphics[width=0.45\textwidth]{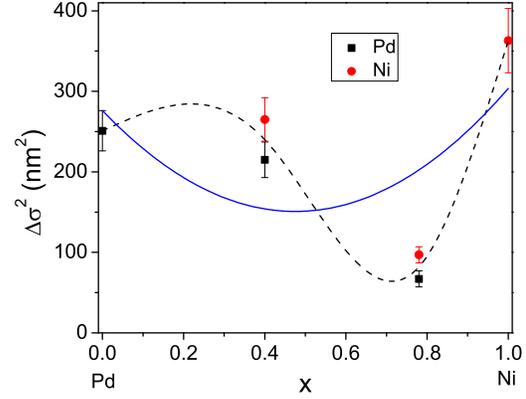}
\caption {Variation of $\Delta \sigma^2$ with $x$ for Pd and Ni
interdiffusion in Si. The dotted line is a guide to the eyes. The
concave upward curve shown as continuous line is the crudest
approximation for this variation.}
 \label{fig5}
\vspace{0.0cm}
\end{figure}
\noindent

The strategy for doing the calculations is as follows: (i) Assume
the validity of TSM. (ii) Use the TTM equations (1) and (2) to see
the evolution of Pd$_{1-x}$Ni$_x$ lattice temperature, at various
radial distances from the ion track, with time. At this stage, use
DFT to compute $C_e$ and $G$ appropriate to be taken as inputs to
the TTM equations (6) and (7). For this, construct
Pd$_{1-x}$Ni$_x$ crystal structures of various compositions close
to the experimental ones and calculate electron densities of
states $g (\epsilon)$. Limit the TSM calculations to bulk
Pd$_{1-x}$Ni$_x$, as Si is common substrate to all the samples and
hence its influence is likely to have no $x$ dependence. (iii) See
if the Pd$_{1-x}$Ni$_x$ lattice melts. Otherwise, record the $x$
variation of an average lattice temperature T$_{\rm {av}}$
achievable in the lattice at a fixed $r$, and then derive a
quantity which could be correlated with $\Delta \sigma^2$. (iv)
See whether the variations of this quantity with $x$ follows any
concave upward trend.

\noindent

\subsection{$g (\epsilon)$, $C_e$ and $G$}

The first step for the calculations is to compute $g (\epsilon)$
using DFT. Figure 8 displays the $g (\epsilon)$ curves for
Pd$_{1-x}$Ni$_x$ ($x$ = 0, 0.25, 0.5, 0.75, 1) alloys. Without
interpreting the densities of states, which themselves are rich of
information like $g (\epsilon_F)$, the underlying orbital
hybridizations, etc. and their variation with $x$, which in turn
may shed light on the electronic and magnetic properties of the
Pd$_{1-x}$Ni$_x$ alloys,\cite{swain2017} we move over to directly
using them as an input to the TTM equations. We restrict the
refinements of the parameters used in the TTM equations only to
the electronic part, which is accessible through DFT. For the rest
of the parameters, like even the phonon part $\lambda \left\langle
\omega^2\right\rangle$ in Eq. (7), we take, as the first
approximation, the terminal ($x$ = 0, 1) values from literature,
and linearly interpolate these values for the intermediate
compositions. The $\lambda \left\langle \omega^2\right\rangle$
values taken for $x$ = 0 and 1 are 41 meV$^2$ and 63 meV$^2$,
respectively \cite{ZW94,DAP76}. Figure 9 (a) displays the
variations of $C_e$ and $G$, respectively, with $T_e$ for the
computed Ni compositions. Instead of investigating how these
parameters vary with $T_e$, finding out how they vary with $x$ for
representative electronic temperatures would be more relevant.
C$_e$($x$) at 5000 K and G($x$) at 300 K, the two temperatures
being representative ones to be used as inputs in the TSM
calculations, are shown in Fig. 9 (b). Thus comes the first
milestone of comparison with the experiment: qualitatively, the
two parameters increase monotonically with $x$ almost together, do
have a shallow concave upward nature, but do not possess a
minimum. We now move forward to performing the TSM
calculations.

\begin{figure}[t]
\vspace{0.0cm} \hspace{0cm}
\includegraphics[width=0.4\textwidth, trim = 00 00 50 40, clip = true]{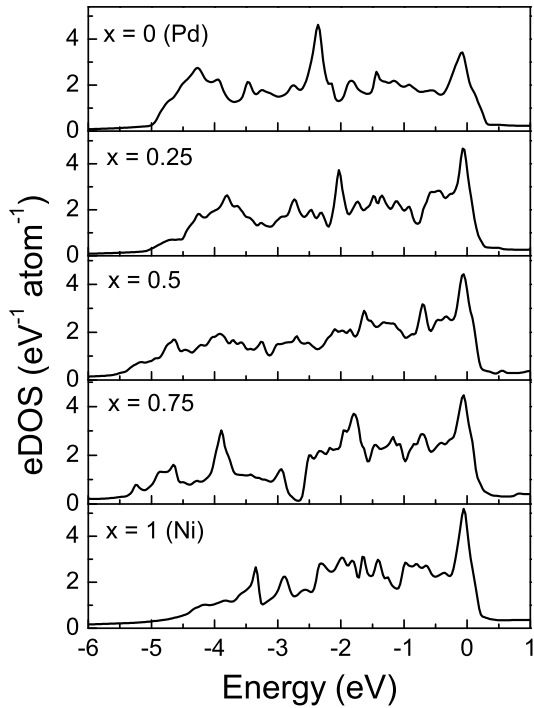}
\caption {The electronic densities of states of the
Pd$_{1-x}$Ni$_x$ alloy system.} \label{fig8} \vspace{0.0cm}
\end{figure}

\begin{figure}%
\centering
\subfigure[][]{%
\label{fig:9-a}%
\includegraphics[width=0.45\textwidth, trim = 00 00 00 00, clip = true]{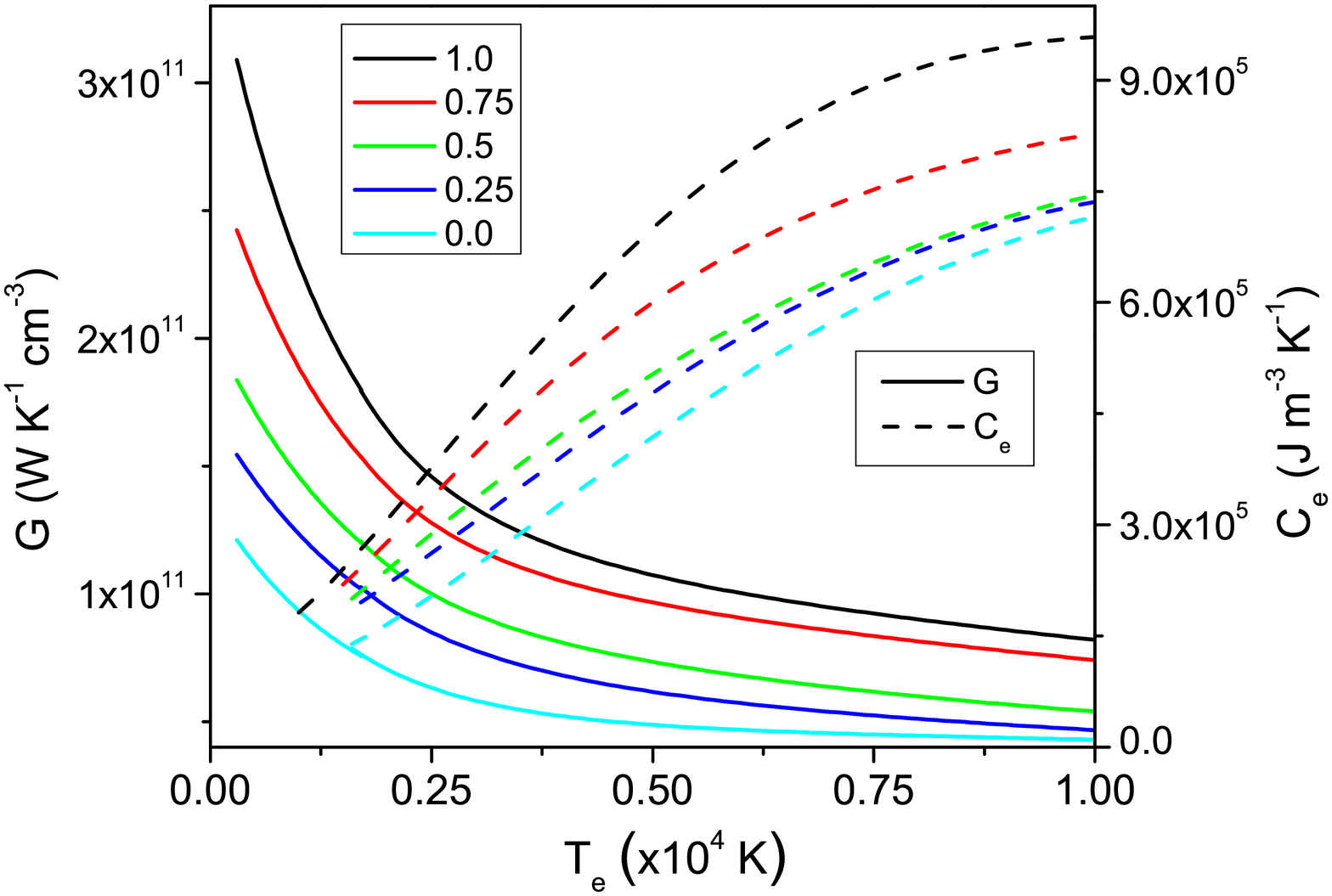}}%
\hspace{8pt}%
\subfigure[][]{%
\label{fig:9-b}%
\includegraphics[width=0.42\textwidth, trim = 20 00 00 00, clip = true]{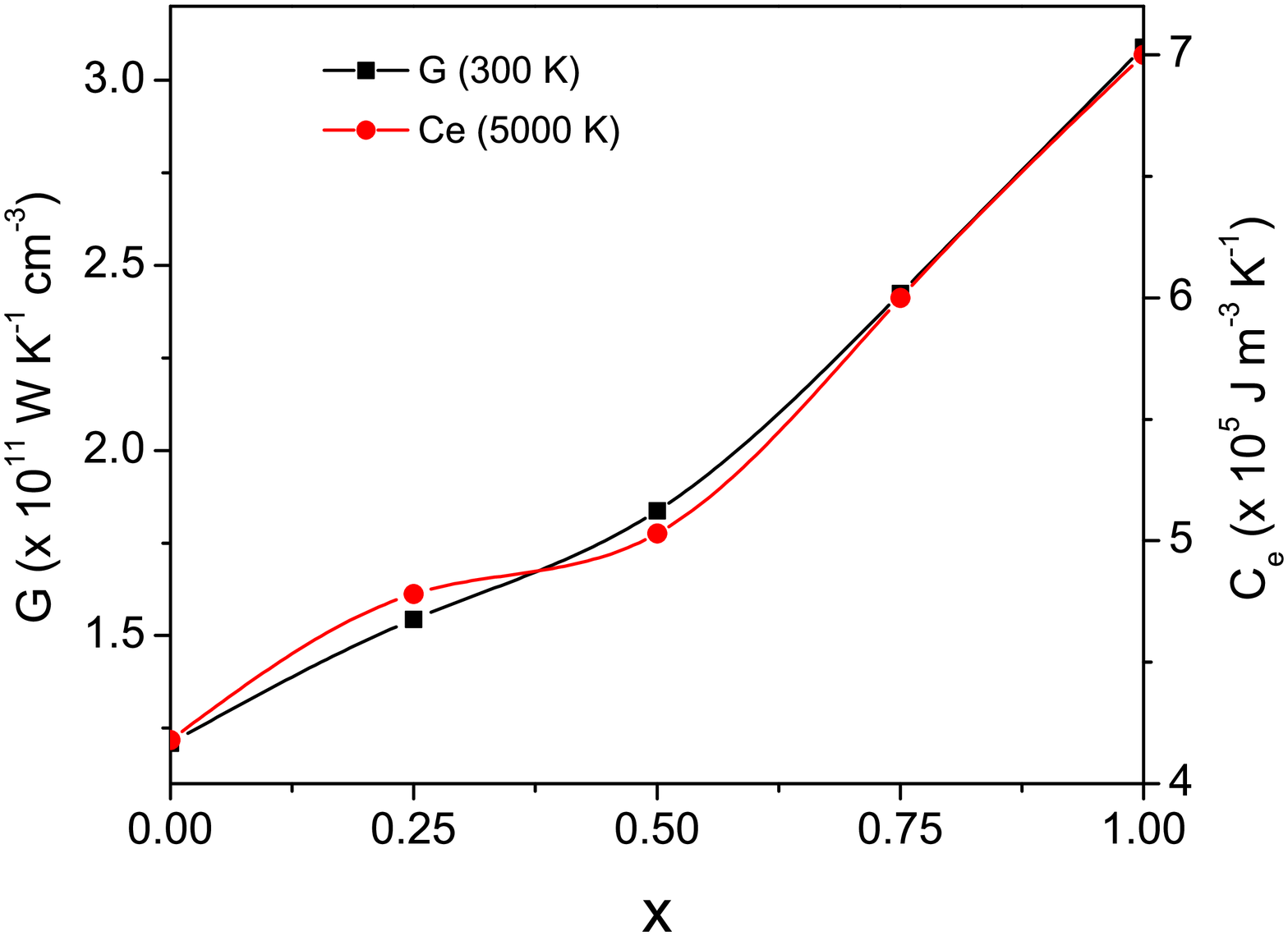}}%
\hspace{8pt}%

\caption{Variations of (a) specific heat (dotted curves) and e-p
coupling strength (continuous lines) with electron temperature,
and (b) of both with Ni composition. The lines in (b) are a guide
to the eyes.} \label{fig9}
\end{figure}

\subsection{TSM calculations}

Table I enlists the additional input parameters that have been
used in the TTM equations (1) and (2) in order to proceed with the
TSM calculations. Figures 10 (a) and (b) show the evolution of
lattice temperature with time at different radial distances from
ion path for the terminal ($x$ = 0 and 1, respectively) systems,
taken as representative ones. As can be observed, the lattice does
not reach its melting temperature\cite{dd} in either case. This
is true for all the compositions, not all shown here though. Thus,
occurrence of molten state diffusion, the requirement for SHI
induced interface mixing according to the TSM, is out of question
in the present case. If the TSM is valid, any possible
interdiffusion must have taken place in the solid state itself
while the lattice was hot, and then would have got enhanced by the
sputtering. To investigate this aspect, we analyse the TSM results
in the following way.

\vspace{0 mm}

\begin{table}

 \caption{\label{}Lattice specific heat (C$_l$), electronic thermal conductivity (K$_e$), lattice thermal conductivity (K) and electronic energy loss (S$_e$), used as inputs to the TSM calculations.}
%\hspace{0mm}
 \begin{tabular}{c c c c c c c}
 \hline
 \hline

\textbf{ Parameter} & \textbf{unit} & \textbf{x = 0} & & \textbf{x = 1} & & \textbf{Ref.}\\
%\textbf{(unit)} & & & & & & \\
 \textbf{(T-range)} & & & & & &\\
%  \hline

 C$_l$ & J/g-K & 0.12 - 0.30 & & 0.27 - 0.60 & &[5] \\
  (90 - 10000 K) & & & & & &\\
%  \hline

  K$_e$  &  W/cm-K  & 0.05 & & 0.05 & & [5]\\
%  (W/cm-K) & & & & & &\\
  (all T) & & & & & &\\
%  \hline

  K$_l$ &  W/cm-K & 2.33 - 0.78 & & 11.33 - 0.50 & & [5]\\
%  (W/cm-K) & & & & & &\\
 (50 - 10000 K) & & & & & &\\
%  \hline

  S$_e$ &  keV/$\rm {\AA}$ & 3.41 & & 3.24 & &[28]\\
%  (keV/$\rm {\AA}$) & & & & & &\\
  (T-independent) & & & & & &\\
  \hline
  \hline

 \end{tabular}

 \end{table}

\begin{figure}%
\centering
\subfigure[][]{%
\label{fig:10-a}%
\includegraphics[width=0.45\textwidth, trim = 00 00 00 00, clip = true]{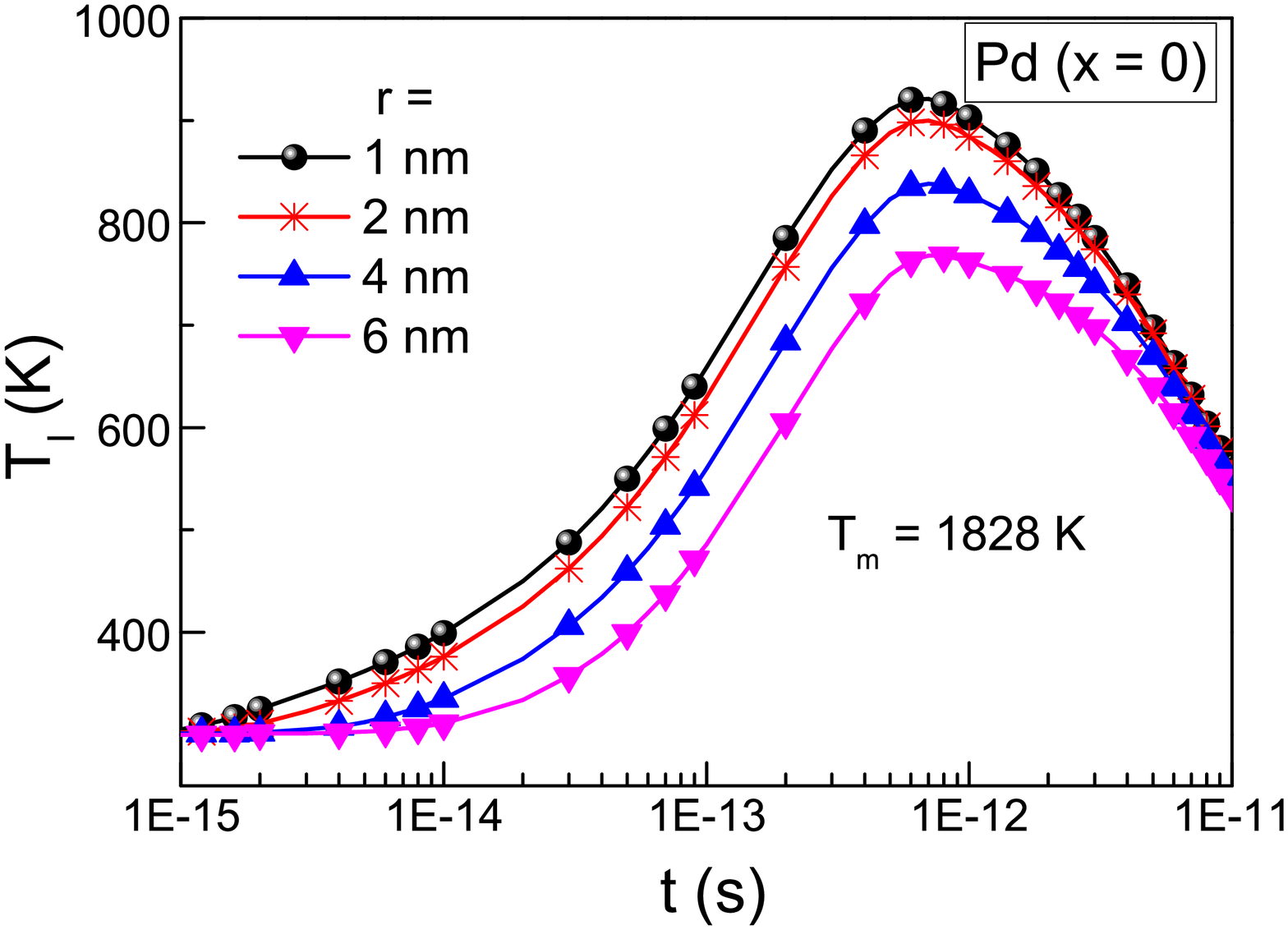}}%
\hspace{8pt}%
\subfigure[][]{%
\label{fig:10-b}%
\includegraphics[width=0.42\textwidth, trim = 20 00 00 00, clip = true]{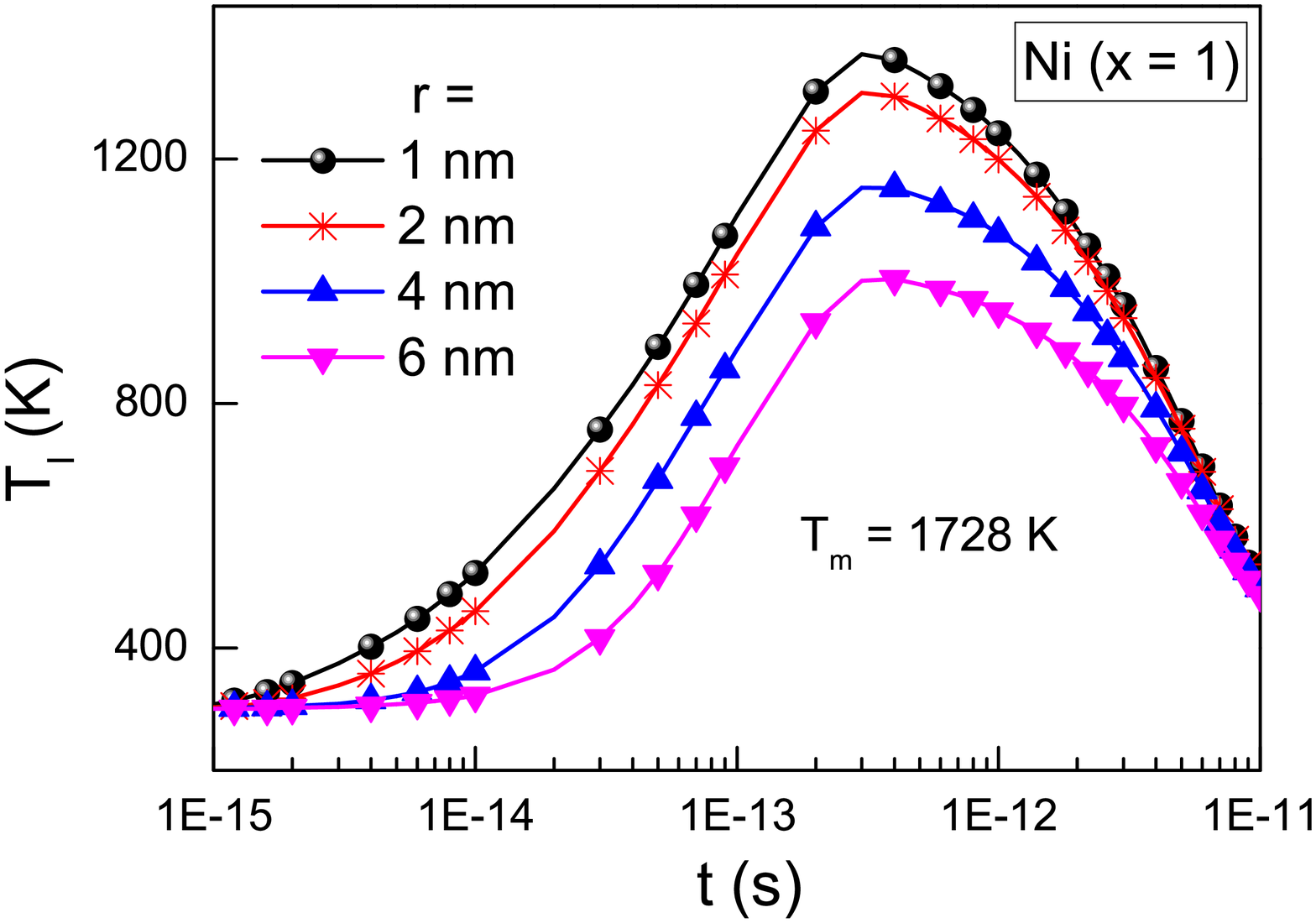}}%
\hspace{8pt}%
\subfigure[][]{%
\label{fig:10-c}%
\includegraphics[width=0.42\textwidth, trim = 20 00 00 00, clip = true]{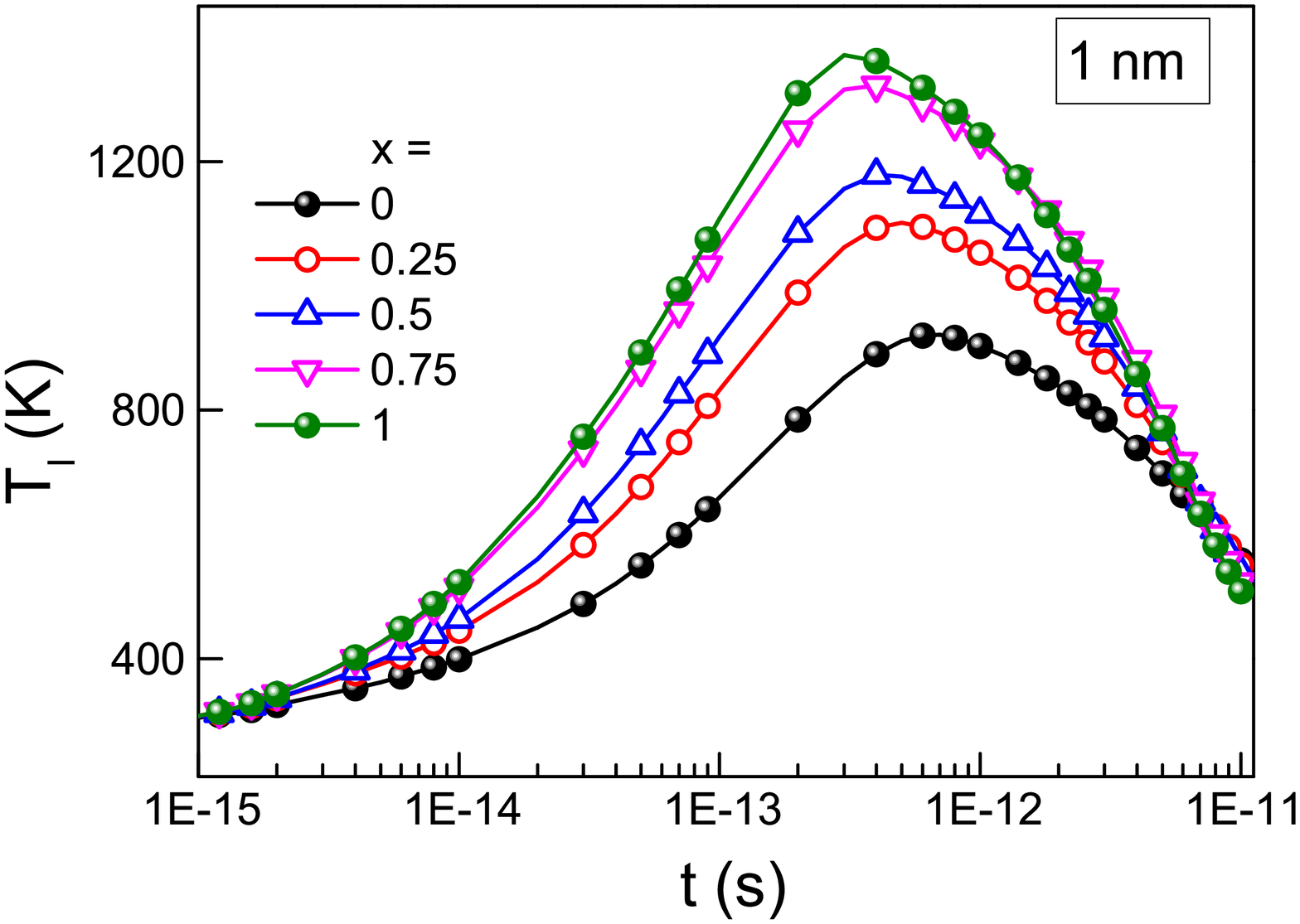}}%
\hspace{8pt}%

\caption{Evolution of lattice temperature with time for (a) $x$ =
0 and (b) $x$ = 1, with radial distance from ion path as a
parameter, and (c) at 1 nm radial distance for all compositions.
Melting temperature T$_{\rm {m}}$ \cite{dd} is indicated.}
\label{fig10}
\end{figure}

Since we are interested in finding out only the $x$ variations of
the derived quantities, we can look at the evolution of lattice
temperature with time, at a common radial distance, say 1 nm, for
all $x$. This is plotted in Fig. 10 (c). As far as diffusion (of
Pd and Ni in Si separately) is concerned, we need to derive an
appropriate average temperature T$_{\rm av}$ at which the
diffusion can be considered to have taken place, and the average
duration $\Delta t$ for the diffusion. A suitable choice for
selecting the two quantities for a particular $x$ would be to take
FWHM of the corresponding T$_l$ versus $t$ curve at 1 nm radial
distance. This sets T$_{\rm av}$ at the temperature where the half
maximum of the curve occurs, while $\Delta t$ is just the FWHM.
Variations of T$_{\rm av}$ and $\Delta t$ with $x$, as derived
from Fig. 10 (c), are shown in Fig. 11 (a). Then the Arrhenius
equation $D = D_0~ {\rm exp} (-E_a/k_{\rm B}~T_{\rm av})$ for
diffusion coefficient $D$ can be used to calculate the square
L$^2$ of the diffusion length as L$^2$ $\propto$ $D \Delta t$,
which itself should be proportional to $\sigma^2$ measured
experimentally, if the TSM is valid. Here, $D_0$ is a
pre-exponential factor, $E_a$ is the activation energy for
diffusion, and $k_{\rm B}$ is the Boltzmann's constant. According
to the literature,\cite{Diff} $D_0$ and $E_a$ for Pd diffusion in
Si are 3.13$\times$10$^{-4}$ cm$^2$/s and 1.1 eV, respectively.
For Ni diffusion in Si, these values are 6$\times$10$^{-4}$
cm$^2$/s and 0.67 eV, respectively.

\begin{figure}%
\centering
\subfigure[][]{%
\label{fig:11-a}%
\includegraphics[width=0.42\textwidth, trim = 00 00 00 00, clip = true]{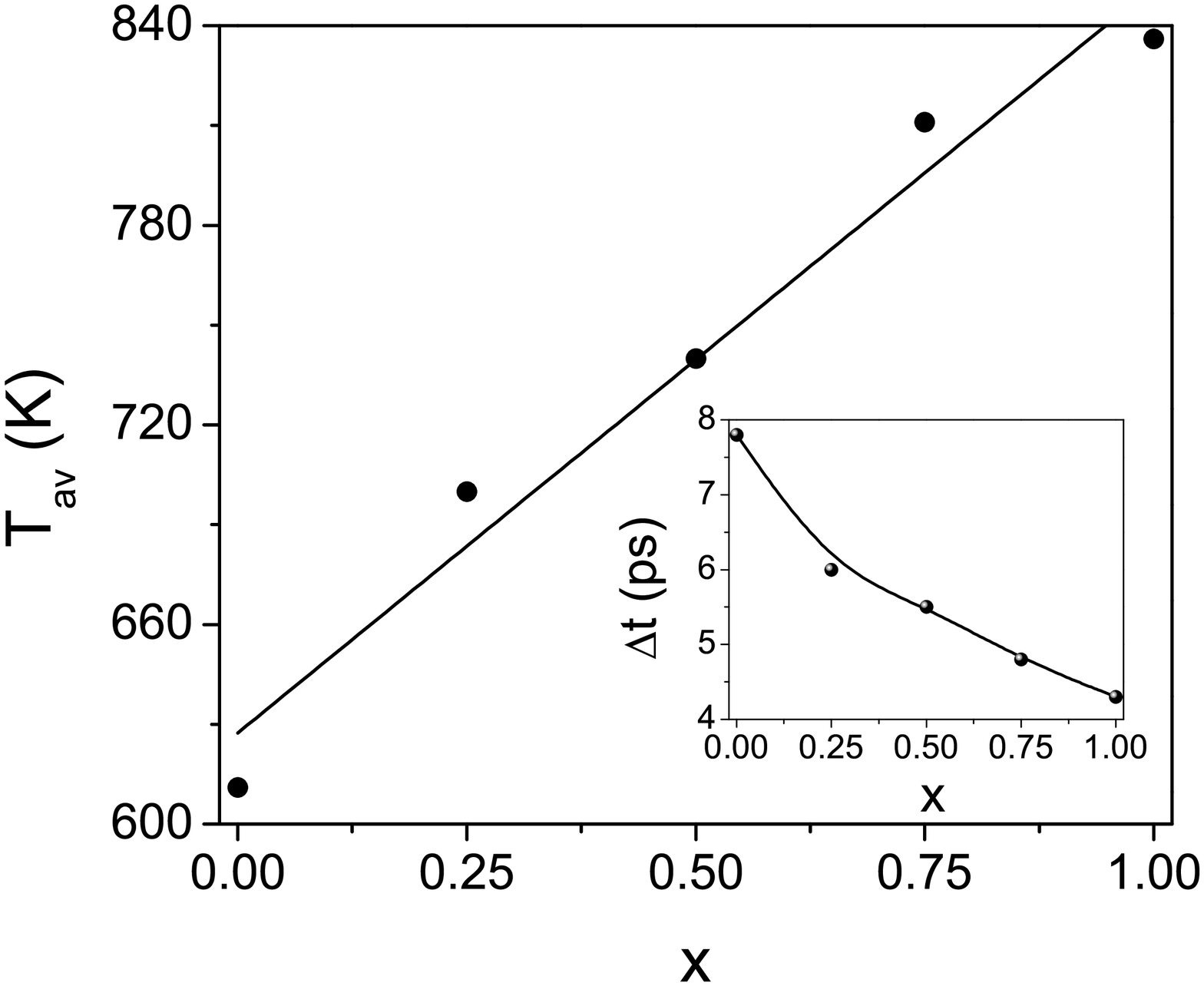}}%
\hspace{8pt}%
\subfigure[][]{%
\label{fig:11-b}%
\includegraphics[width=0.42\textwidth, trim = 00 00 00 00, clip = true]{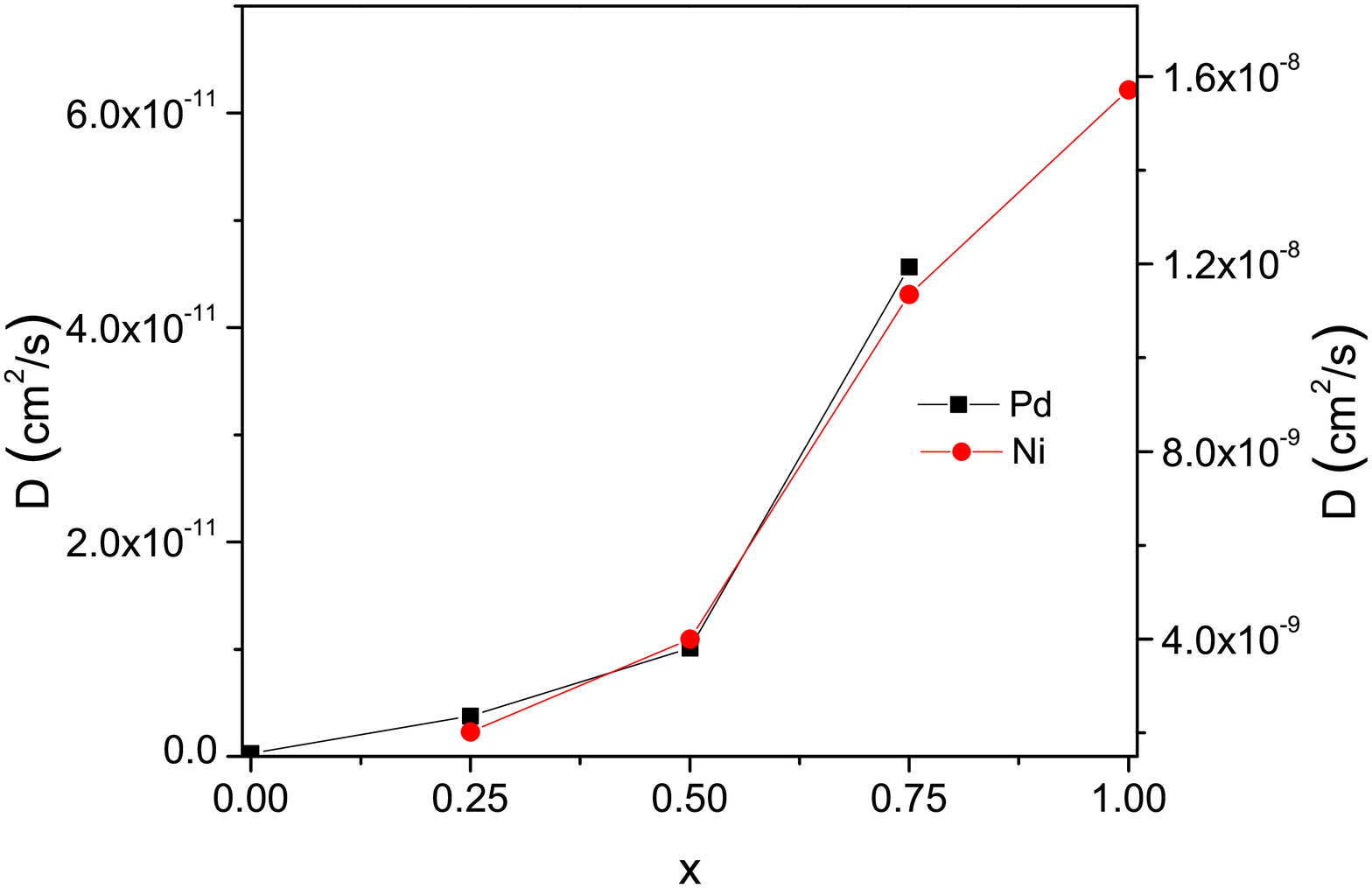}}%
\hspace{8pt}%
\subfigure[][]{%
\label{fig:11-c}%
\includegraphics[width=0.42\textwidth, trim = 00 00 00 00, clip = true]{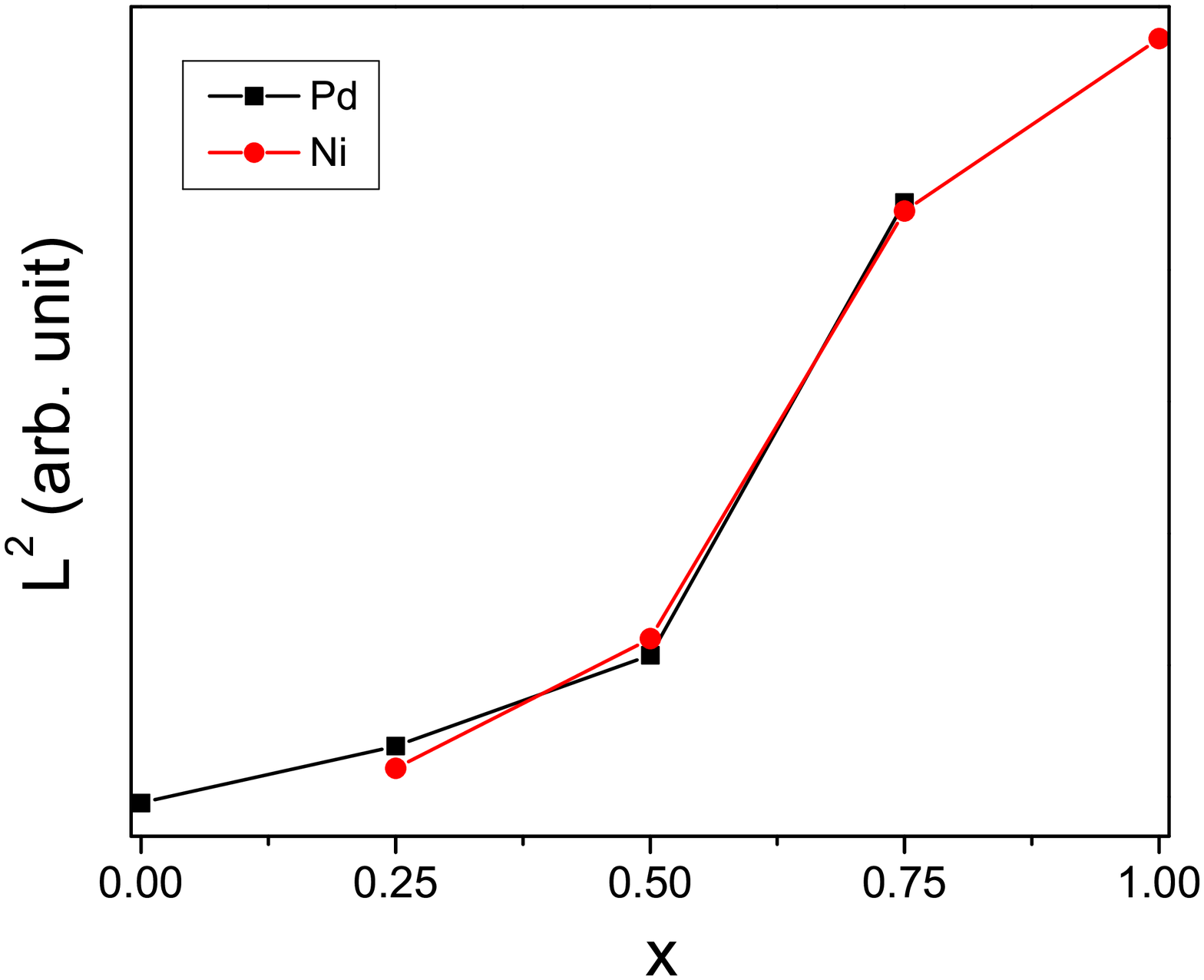}}%
\hspace{8pt}%

\caption{(a) Variations of average lattice temperature T$_{\rm
av}$ at 1 nm radial distance from ion path, and of the duration
$\Delta t$ (inset) of this temperature with $x$. (b) Pd and Ni
diffusion coefficients at these temperatures as a function of $x$.
(c) L$^2$ versus $x$ curves for Pd and Ni. Lines are a guide to
the eyes.} \label{fig11}
\end{figure}

Plots of the diffusion coefficients thus calculated are shown in
Fig. 11 (b). It can readily be inferred from the figure that
although the two diffusion coefficients follow the same $x$
variation, they are three orders of magnitude apart. Figure 7,
however, suggest that if the experimentally observed change in
interfacial width on irradiation is due to interdiffusion, both
the diffusion coefficients, or at least the effective diffusion
coefficients, must be almost equal. So, the experimental
observations can not be a result of simple solid state diffusion.
Liquid (molten) state diffusion has already been ruled out. In
this scenario, only one proposition seems to work: SHI irradiation
must have produced an enormous amount of defects, thereby reducing
the activation energy for diffusion to such an extent that the
effective diffusion coefficients are liquid-like. The liquid state
diffusivities are of the same order of magnitude (10$^{-5}$
cm$^2$/s).\cite{SKS05,Liu12,Pradhan71} Once the activation energy
is reduced, even the Ar$^+$ ion sputtering must be able to produce
the observed diffusion. So, we ignore the magnitudes of the
diffusion coefficients and focus only on their variation with $x$.
We can see, from Fig. 11 (b), that this variation is monotonic,
and certainly does not follow a concave upward trend. This can be
considered as the second milestone for comparison, where
experimental results do not seem to be reproduced by theory. Next,
since $\Delta t$ changes only by a factor of two from Pd to Ni,
L$^2$ must also follow the same trend. The variation of L$^2$ with
$x$ determined this way has been plotted in Fig. 11 (c) for
interdiffusion of both Pd and Ni in Si. As can be seen, this no
way resembles a concave upward curve with a minimum. The last thing to check is the influence of $x$-dependence of crystallite size, based on the reports,\cite{Berthe98,Berthe2000} on nanometric size effects on irradiation induced changes. The crystallite sizes of all the samples as determined using the Debye-Scherrer equation\cite{Patt39} applied to the XRD patterns (Fig. 1) range from 20 nm to 60 nm. This size variation hardly changes the influence of SHI's, as can be seen in the references 50 - 51. So, at last, it seems that the TSM, which has so far been successful in explaining the experimental outcomes of SHI - matter interaction, is not able to explain the experimental results of SHI mixing presented in this work. Perhaps some other considerations, apart from the TSM suggested molten state diffusion for SHI mixing, need to be explored.

\section{Conclusion}

The thermal spike model of SHI - matter interaction has been
assessed via its applicability in SHI mixing by comparing the
results of (i) 100 MeV Au ion irradiation of Pd$_{1-x}$Ni$_x$ thin
films deposited on Si, and (ii) calculations based on density
functional theory and the model. The key concept behind the
calculations is the well accepted notion that SHI mixing is a
result of diffusion in transient molten state created by the SHI.
Although no mixing has been detected in RBS spectra, a
considerable amount is observable on Ar$^+$ ion sputtering based
depth profiling associated with XPS characterizations. This is
proposed to be possible due to both the high spatial resolution
obtainable from XPS and the sputtering assisted mixing. Since the
latter has to be uniform in all the samples, its role can be
ignored in the study of $x$ dependence of mixing, which has been
the prime focus of the work. The irradiation induced change in the
variance of the depth profile determined from the XPS spectra
follows an $x$ variation which is concave upward with a pronounced
minimum. DFT has been used to derive the $x$ dependences of
electronic specific heat and electron-phonon coupling strength.
These data have been used as an input to the TTM equations
appropriate to the thermal spike model to calculate the evolution
of lattice temperature with time at various radial distances from
the ion path. Finally, a quantity L$^2$ proportional to the
variance is calculated, and its variation with $x$ is derived. If
the TSM, along with the conception that mixing is a result of
molten state diffusion is valid, L$^2$ must also vary with $x$
along a curve which is concave upward with minimum. However, L$^2$
increases monotonically with $x$ without any minimum, and thus is
not in accordance with the experimental observations. Even the crystallite size variation, as determined from the XRD patterns, can not account for the observed $x$-variation of $\sigma^2$. This leads to the conjecture that the combination of the TSM and the molten state diffusion is probably an insufficient description of SHI mixing, and hence the underlying mechanism requires further considerations.

\vspace{10 mm}

\section {Acknowledgements}
\vspace{10 mm}

D. Kabiraj and S. R. Abhilash of the Target Lab of IUAC, New Delhi are acknowledged for their help in sample preparation. The help provided by S. Ojha and G. R. Umapathy of PARAS facility of IUAC, New Delhi in RBS measurements is also acknowledged. The authors also acknowledge the support by the Pelletron Accelerator Group of IUAC, New Delhi for irradiations. Many useful discussions with M. Toulemonde are also highly acknowledged. Paramita Patra is thankful to Indian Institute of Technology Kharagpur for the financial support to carry out the research work. 

\vspace{10 mm}

%{\bf References}

\end{document}